\documentclass[aps,prb,twocolumn,notitlepage,showpacs,floatfix,superscriptaddress]{revtex4-2}

\usepackage{bm}
\usepackage{amsmath,amssymb,graphicx}
\usepackage{bbold}
\usepackage[titletoc]{appendix}
\usepackage[colorlinks=true,citecolor=blue,urlcolor=blue,linkcolor=blue]{hyperref}
\usepackage{braket,titlesec,natbib}
\usepackage{dsfont}
\usepackage{comment}
\usepackage{float}
\usepackage{color}
\usepackage{hyperref}
\usepackage{graphicx}
\usepackage{soul}
\usepackage{ulem}

\definecolor{darkgreen}{rgb}{0.0, 0.42, 0.10}

\newcommand{\bse}{\begin{subequations}}
\newcommand{\ese}{\end{subequations}}

\newcommand{\bs}{\boldsymbol{\sigma}}

\newcommand{\beq}{\begin{equation}}
\newcommand{\eeq}{\end{equation}}
\newcommand{\bea}{\begin{eqnarray}}
\newcommand{\eea}{\end{eqnarray}}
\newcommand{\ve}{\varepsilon}
\newcommand{\up}{\uparrow}
\newcommand{\down}{\downarrow}

\newcommand{\bk}{{\bf k}}

\newcommand{\bq}{{\bf q}}

\newcommand{\bu}{{\bf u}}
\newcommand{\bP}{{\bf P}}

\newcommand{\bwt}{\begin{widetext}}
\newcommand{\ewt}{\end{widetext}}

\newcommand{\er}{\eqref}

\begin{document}

\title{Phonon-Induced Collective Modes in Spin-Orbit Coupled Polar Metals}

\author{Abhishek Kumar}
\affiliation{Department of Physics and Astronomy, Rutgers University, Piscataway, New Jersey 08854, USA}
\affiliation{Département de physique and Institut quantique, Université de Sherbrooke, Sherbrooke, Québec, Canada J1K 2R1}
\author{Premala Chandra}
\affiliation{Department of Physics and Astronomy, Rutgers University, Piscataway, New Jersey 08854, USA}
\author{Pavel A. Volkov}
\affiliation{Department of Physics and Astronomy, Rutgers University, Piscataway, New Jersey 08854, USA}
\affiliation{Department of Physics, Harvard University, Cambridge, Massachusetts, 02138 USA}
\affiliation{Department of Physics, University of Connecticut, Storrs, Connecticut 06269, USA}

\date{\today}
\begin{abstract}
	We study the interplay between collective electronic and lattice modes in polar metals in an applied magnetic field aligned with  the polar axis. Static spin-orbit coupling leads to the appearance of a particle-hole spin-flip continuum that is gapped at low energies in a finite field. We find that a weak spin-orbit assisted coupling between electrons and polar phonons leads to the appearance of new electronic collective modes. The strength of the applied magnetic field tunes the number of modes and their energies, which can lie both above and below the particle-hole continuum. For a range of field values, we identify  Fano-like interference between the electronic continuum and phonons. We show that signatures of these collective modes can be observed in optical spectroscopy experiments, and we provide the corresponding theoretical predictions.
\end{abstract}

\maketitle
\section{Introduction}

Polar metals are metallic analogs of ferroelectrics,
as in these materials, a transition isostructural to a polar one occurs without the development of a macroscopic electric dipole moment; these systems break inversion symmetry 
\cite{Anderson:1965, shi:2013, Fei:2018, Cao:2018, jin:2019,laurita:2019,Zhou_review}.
Polar metals have recently been proposed as promising platforms for the realization of strongly correlated electronic phases, particularly near polar quantum critical points  \cite{Fu:2015, kozii:2015,ruhman2017,kozii:2019,volkov2021,kozii2022,klein2022}. These studies have been motivated by the unconventional transport and superconducting properties of SrTiO$_3$; consensus on the origin of its superconductivity has not yet been achieved, and most theoretical proposals rely on novel forms of electron-phonon coupling \cite{edge2015, Ruhman:2016, Maria:review, Maria:2020,kanasugi:2019,feigelman:SC,volkov2021}.

One class of exotic pairing proposals \cite{Fu:2015, kozii:2015} involves the coupling of the polar displacement to the electronic spin current; it may be particularly relevant for KTaO$_3$ \cite{ktoSCexp,Venditti_2023}.  Recently, we have shown \cite{kumar_polar} that the strength of this spin-orbit mediated electron-phonon interaction in a nearly polar metal can be measured by probing the evolution of its collective excitations in an applied magnetic field;
the polar phonon hybridizes with the Silin-Leggett modes \cite{silin1958, baym, statphys, nozieres}, resulting in an avoided
crossing and a gap that is a function of the coupling strength \cite{kumar_polar}.

Here we demonstrate that in a polar metal with spontaneously broken inversion symmetry, new collective modes emerge due to the same Rashba-like electron-phonon coupling mechanism. The breaking of inversion symmetry then
violates Kramers' theorem, lifting the spin degeneracy via Rashba spin-orbit coupling. Previously, electron-electron interactions have been predicted to result in new collective modes in purely electronic systems with Rasbha coupling: chiral spin waves that are oscillations of the magnetization even in the absence of an external magnetic field \cite{maslov:review, shekhter2005, ashrafi2012, zhang2013, ashrafi_2013, maiti2014, kumar:gr, maiti2016, kumar2017, raines:2022}. 
These chiral spin waves have been subsequently  observed in Raman spectroscopy in several systems with \cite{perez:2013, perez:2015, perez:2016, karimi:2017} and without \cite{kung2017} external magnetic fields. However the possibility of new collective excitations due to spin-orbit assisted electron-phonon interactions, particularly with low-energy phonons, has not been addressed in these previous reports.

\begin{figure}[h]
 \centering
 \includegraphics[scale=0.59]{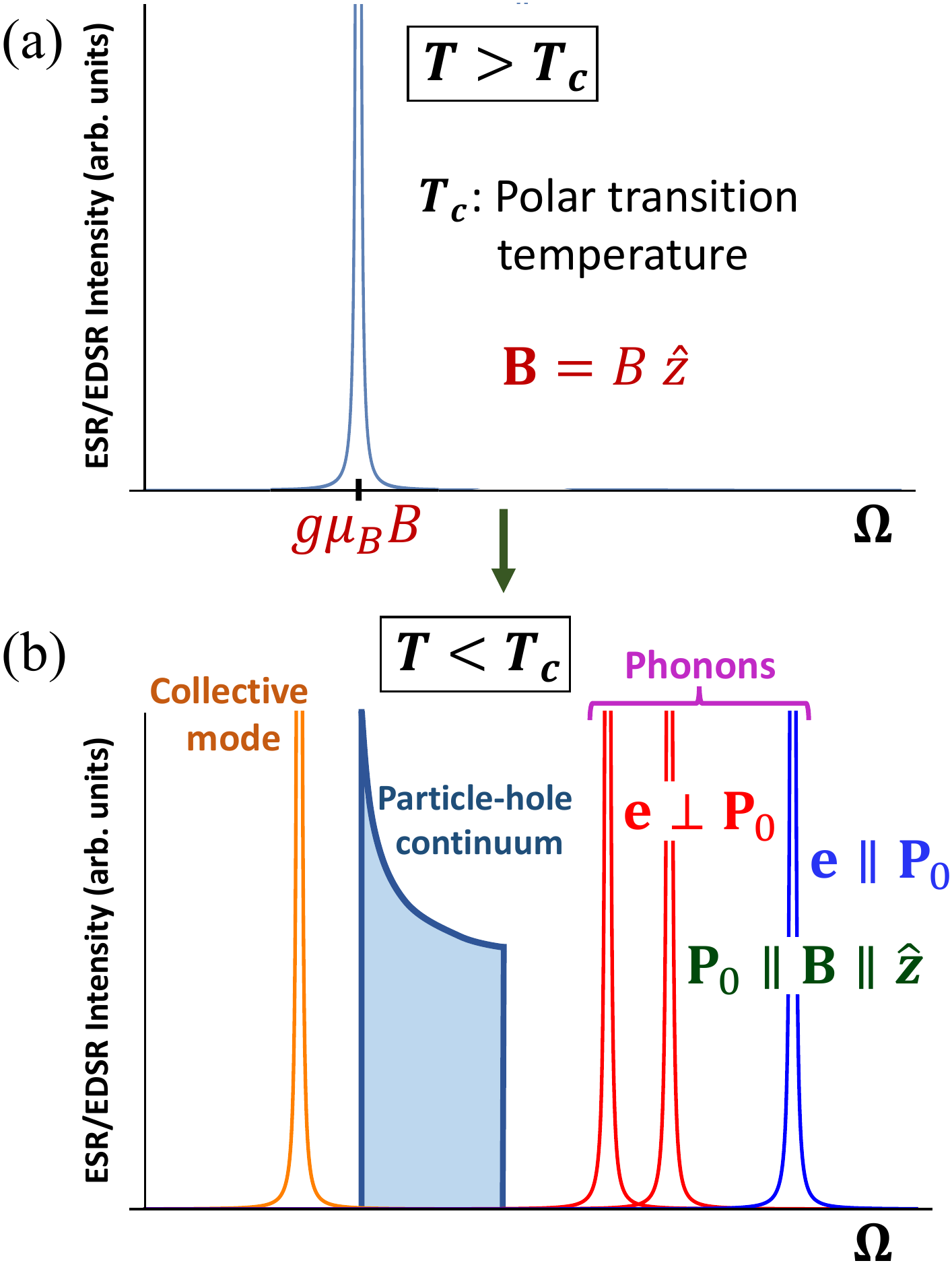}
 \caption{\label{fig:Intro} 
 A schematic of the predicted spin resonance response, electron spin resonance (ESR) and electric-dipole spin resonance (EDSR), of collective modes in a weakly spin polarized metal 
 in the (a) Non-Polar (b) Polar Phases. 
 (a) In the non-polar phase ($T>T_c$), the only spin response signal is due to the spin-flip transition at the Zeeman energy. 
 (b) In the polar phase ($T<T_c$), several new features appear in the spin response spectrum:
 There is a gapped particle-hole continuum 
 (blue shaded region), and a finite response from polar phonons polarized parallel (blue) and perpendicular (red) to the polar axis. Finally, the spin-orbit assisted electron-phonon interaction leads to the appearance of an electronic collective mode below the continuum (yellow). The energy and number of these modes can be controlled by the magnitude of the magnetic field as discussed in the text.}
\end{figure}

In the present paper we consider the emergence of collective modes in a polar metal close to its polar transition that is driven by phonon softening. 
We find that the spin-splitting due to the emergent violation of Kramers' theorem
results in the appearance of electronic collective modes and phonon Fano resonances even in zero magnetic fields due to the presence of a structured spin excitation continuum. At zero fields, the spin-excitation continuum is gapless at low energies, and
interaction-induced modes
exist only above the spin excitation continuum.
However an applied magnetic field along the polar axis gaps out the continuum at low energies
and collective modes emerge both below the gap and above the gap even at weak coupling. 
Tuning of the applied field strength and proximity to the polar transition allows one to control the energy, number and character (electronic or phononic) of these collective modes. Finally, we show that these modes can be observed as resonances in electronic optical spectroscopy (electron spin resonance (ESR) and electric-dipole spin resonance (EDSR)) experiments, 
as illustrated schematically in Fig.~\ref{fig:Intro}; both ESR and EDSR show resonances at same frequencies. 
These spectroscopies probe only spin-flip excitations. However, the difference is that in the non-polar phase ($T>T_c$) 
it has a sole feature at the Zeeman energy (Fig.~\ref{fig:Intro}(a)) which is protected from renormalization due to the spin-orbit assisted electron-phonon interaction, whereas in the polar phase ($T<T_c$), ESR and EDSR capture interaction effects and probe interaction-induced renormalized phonons (red and blue peaks) as well as electronic collective modes (yellow peak) both above and below the particle-hole continuum (Fig.~\ref{fig:Intro}(b)). We note that these phenomenona can occur near both classical and quantum phase transitions ($T_c \to 0$), where in the latter case the tuning parameter could be non-thermal with examples include pressure or chemical doping \cite{rowley,Chandra:2017}.
Our work demonstrates the emergence of magnetically active collective modes arising from interaction between electrons and  the crystalline lattice fluctuations.

The rest of the paper is organized as follows. In Sec.~\ref{model} we describe our model and our general strategy of studying the collective modes and the electronic spin response in polar metals with spin-orbit coupling and Zeeman field. First, in Sec.~\ref{GL} we adopt a Ginzburg-Landau approach to derive the phonon propagator in the polar phase, and then in Sec.~\ref{RPA} discuss spin susceptibility calculation method. In the same Sec.~\ref{RPA} we also provide a relation between ESR and EDSR responses which indicate that the resonance features in both show up at same frequencies. Sec.~\ref{coll_mod} deals with the spectra of collective modes in polar metals. In Sec.~\ref{spin-current_tensor}, we discuss phonon self-energy at zero (Sec.~\ref{scsc_Z=0}) and finite (Sec.~\ref{sec:se_Z}) magnetic fields. In Sec.~\ref{CM} we discuss phonon response, or poles of full phonon Green's function (collective modes), at zero (Sec.~\ref{sec:zero_field}) and finite (Sec.~\ref{finite_field}) magnetic fields. Sec.~\ref{ESR} is devoted to the details of the electronic spin response where we discuss the excitation of collective modes at zero (Sec.~\ref{ESR_zero}) and finite (Sec.~\ref{ESR_finite}) magnetic fields. Finally, in Sec.~\ref{dis} we present our conclusions and discuss the experimental prospects of the effects predicted in this paper. Technical details of the calculations are delegated to Appendices \ref{appen:coh} and \ref{appen:bubble}.

\section{Model and General Formalism} 
\label{model}
In this section we present our general strategy to study phonon-induced electronic collective modes in spin-orbit coupled polar metals under applied magnetic field. Collective modes can be identified with the poles of the phonon propagator renormalized  by the electron-phonon interaction \cite{kumar_polar}. 
Furthermore, these collective modes can be accessible to ESR and EDSR experiments if they result in poles of their corresponding response functions, spin susceptibility and optical conductivity.
This requires broken inversion symmetry so these measurements must be performed in the polar phase, as will be discussed in detail in Sec. \ref{ESR}.

We now a construct a Hamiltonian for an ordered polar metal with spin-orbit assisted electron-phonon interactions in an applied
magnetic field. We start with the electronic part, and the phonon contribution will be discussed in Sec.~\ref{GL}.

Let us consider a single parabolic band model for the conduction electrons.
We also assume that $\Delta_Z \ll \mu$, where $\mu$ and $\Delta_Z = g \mu_B B$ are the chemical potential and the Zeeman splitting respectively;   here $\mu_B $ is the Bohr magneton,  $g$ is the Lande $g$-factor and $B$ is the strength of the applied
magnetic field. 
Without loss of generality we take the local polar order to be aligned along the $\hat{z}$-axis. 
The full electronic Hamiltonian can then be written as 
\beq
\hat H = \hat{H}_0+\hat{H}_\text{el-ph},
\eeq
where $\hat{H}_0$ is given by
\beq
\label{single1}
\hat{H}_0 = \sum_{\bk, s} \ve_\bk \psi_{\bk, s}^\dagger \psi_{\bk, s}  + \frac{1}{2} g \mu_B \sum_{\bk, s, s'} \textbf{B} \cdot \hat{\bs}_{ss'} \psi_{\bk, s}^\dagger \psi_{\bk, s'}.
\eeq
Here $\ve_\bk = k^2/2m-\mu$ is the single electron dispersion, $m$ is the band mass, and $\textbf{B} = B \hat{z}$ is the magnetic field, chosen to be parallel to the polar axis. 
As we will discuss shortly in Sec.~\ref{sec:se_Z}, only this orientation leads to the opening of a gap in the spin excitation continuum for weak fields, resulting in the new collective modes.
We note that here we do not consider orbital magnetism, since ultimately we are interested in identifying experimental signatures of the collective modes we are studying. At finite temperatures with everpresent disorder, orbital magnetism
is suppressed strongly compared to its spin counterpart; this phenomenon is known as the "Dingle reduction factor" and more 
details can be found in Ref.~\cite{kumar_polar}.

For $\hat{H}_\text{el-ph}$, we consider a spin-orbit assisted electron-phonon coupling given by a Rashba-type Hamiltonian \cite{Fu:2015, kozii:2015, Ruhman:2016, kanasugi:2018, kanasugi:2019, kozii:2019, Maria:review, Maria:2020, kumar_polar, Maria_polar, efrat:2023}:
	\beq
	\label{coupling}
	\hat{H}_\text{el-ph} = \lambda \sum_{\bk, \bq} \sum_{s, s'} \psi_{\bk + \bq/2, s}^\dagger \big[ (\bk \times \hat{\bs}_{ss'}) \cdot \bP_{\bq} \big] \psi_{\bk - \bq/2, s'}
	\eeq
where $\lambda$ is the 
electron-phonon coupling constant, $\psi_{\bk, s}^\dagger (\psi_{\bk, s})$ is the electron creation (annihilation) operator with momentum $\bk$ and spin $s = \up, \down$, $\hat{\bs}$ is the Pauli matrix for spin (or Kramers ``pseudospin" quantum number) and $\bP_\bq$ is the local polarization of the crystal at finite momentum $\bq$ which is proportional to the optical phonon displacement
field $(\bu_\bq)$, ionic charge density $(n_0)$ and Born effective charge $(Ze)$ according to the formula $\bP_\bq = n_0 (Ze) \bu_\bq$.
In the polar metal phase, the local polarization field $\bP_\bq$ can be expressed as a sum of two terms,
\beq
\label{pol}
\bP_\bq = \bP_0 \delta(\bq) + \delta\bP_\bq,
\eeq
where $\bP_0$ is a finite expectation value at $q=0$ and
$\delta\bP_\bq$ refers to the fluctuations
about this average value. Thus, $\hat{H}_\text{el-ph}$ can be split in two parts:
\begin{equation}
\hat{H}_\text{el-ph} =  \hat{H}_\text{SOC} + \delta
\hat{H}_\text{SOC}.
\end{equation}
The second term, $\delta
\hat{H}_\text{SOC}$, is simply equivalent to \eqref{coupling} with $\bP_\bq \to \delta\bP_\bq$.
$\hat{H}_\text{SOC}$, on the other hand is obtained by substituting the first term of ~\er{pol} into ~\er{coupling}. One obtains then the (static) Rashba SOC Hamiltonian \cite{rashba1960, bychkov1984}:
\beq
\label{stat_rash}
\hat{H}_\text{SOC} = \alpha \sum_{\bk, s, s'} \psi_\bk [(\bk \times \hat{\bs}_{ss'}) \cdot \hat{n}] \psi_\bk,
\eeq
where $\hat{n}$ is a unit vector parallel to the polar order parameter $\bP_0$, with $\bP_0 \parallel \hat{z}$, and $\alpha = \lambda |\bP_0|$.
We emphasize that this static Rasbha interaction \er{stat_rash} is absent in the non-polar phase \cite{kumar_polar}.

\begin{figure}[h]
	\centering	\includegraphics[scale=0.58]{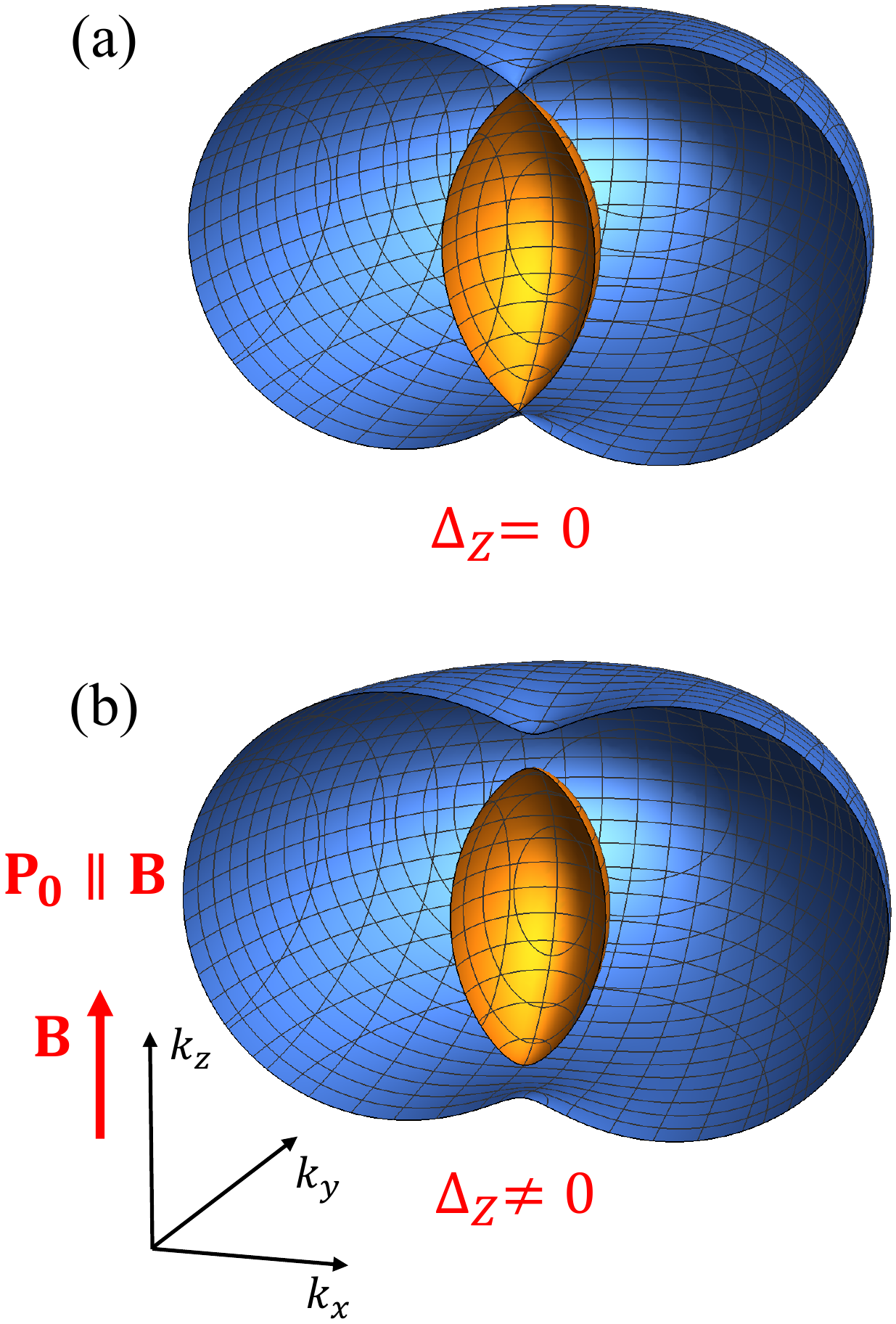}
	\caption{\label{fig:dis} Schematic of the Fermi surface in the presence of a static  
 Rashba spin-orbit interaction (a) in the absence and (b) in the presence of a magnetic field parallel to the polar axis ($\bP_0 \parallel \hat{z}$). The colors mark the Fermi surfaces of the two spin-split subbands. The Zeeman splitting (b) removes the degeneracy between bands along the $z$ axis and leads to a gap in the spin excitation spectrum.}
\end{figure}

The eigenvalues and eigenvectors of the single particle Hamiltonian
\beq
\label{single}
\hat{H}_\text{single} = \hat{H}_0 + \hat{H}_{\text{SOC}},
\eeq
defined in Eqs.~\er{single1} and \er{stat_rash} are
\beq
\label{es}
\begin{split}
\ve_\bk^r &= \frac{k^2}{2m_b} + \frac{r}{2} \Delta_\bk, ~\Delta_\bk = \sqrt{4\alpha^2k^2 \sin^2\theta + \Delta_Z^2}; \\
\ket{r} &=
\left( \begin{array}{c}
\frac{i r e^{-i \phi} (r\Delta_Z + \Delta_\bk)^{1/2}}{\sqrt{2\Delta_\bk}} \\
\frac{(-r\Delta_Z + \Delta_\bk)^{1/2}}{\sqrt{2 \Delta_\bk}}
\end{array} \right)
\end{split}
\eeq
where $r=\pm1$ is the chirality of the spin-split subbands, and $\theta$ and $\phi$ are the polar and azimuthal angles of $\bk$, respectively. At finite doping, the contour of two static-Rashba SOC split Fermi surfaces in the absence and in the presence of an applied magnetic field is shown in Figs.~\ref{fig:dis}(a) and (b), respectively. 

The Hamiltonian \er{single} does not include Coulomb interactions between electrons. Indeed, in polar metals near
polar criticality, Coulomb interactions effects on electrons are suppressed by the large dielectric constant \cite{kumar:2021, kumar_polar, volkov2021}. Furthermore, sufficient density of conduction electrons make the effects of Coulomb interaction on phonons (i.e. the splitting between  longitudinal optical (LO) phonons and transverse optical (TO) ones) negligible too \cite{kumar_polar}. Here we will always assume carrier densities to be large enough such that the LO-TO distinction can be neglected.

\subsection{Anisotropy in the Phonon Propagator} 
\label{GL}
We use a symmetry-based approach to determine the form of the bare phonon propagator in the ordered polar metallic phase. The Matsubara (imaginary) time Lagrangian density in the phonon field for cubic symmetry up to the quartic order is given by \cite{roussev:2003}
\beq
\label{fe}
{\mathcal L}_{ph}  =
\frac{2\pi}{\Omega_0^2}
\left(
\frac{1}{2}\sum_i \dot{P}_i^2 + \frac{w^2}{2} \sum_{i} P_i^2 + \frac{1}{4} \sum_{ij} (g + v_i \delta_{ij}) P_i^2 P_j^2
\right),
\eeq
where $\dot{P}_i$ is the time derivative of the order parameter, $\delta_{ij}$ is the Kronecker delta, and $i$ and $j$ are the Cartesian indices $x, y$ and $z$. 
The prefactor $2\pi/\Omega_0^2$ is for proper normalization, where
$\Omega_0$ is proportional to the ionic plasma frequency $\omega_{pi}$ \cite{volkov2021,kumar_polar}. In this work we focus on the $q=0$ modes and thus we have omitted spatial derivative terms in the Lagrangian \er{fe}. The quadratic term ($w^2$) of \er{fe} results in the mass (frequency) of the bosons while the quartic terms ($g$ and $v_i$) represent local anharmonic interactions. Here we assume cubic symmetry which allows us to take $v_i = v$. The term proportional to $g$ is rotationally invariant and is thus insensitive to the order parameter orientation. By contrast the term proportional to $v$ breaks the O(3) rotational symmetry down to a discrete one for the cubic lattice. Its sign determines the polarization orientation in the ordered phase (along the main axis or diagonals of the unit cell). For dimensional consistency, we recall that here the order parameter $\bP$ has the same dimension as the electric field. One can then deduce the dimensions of $g$ and $v$ as that of the inverse density of states (in three dimensions).

\begin{figure}
\centering
\includegraphics[scale=0.52]{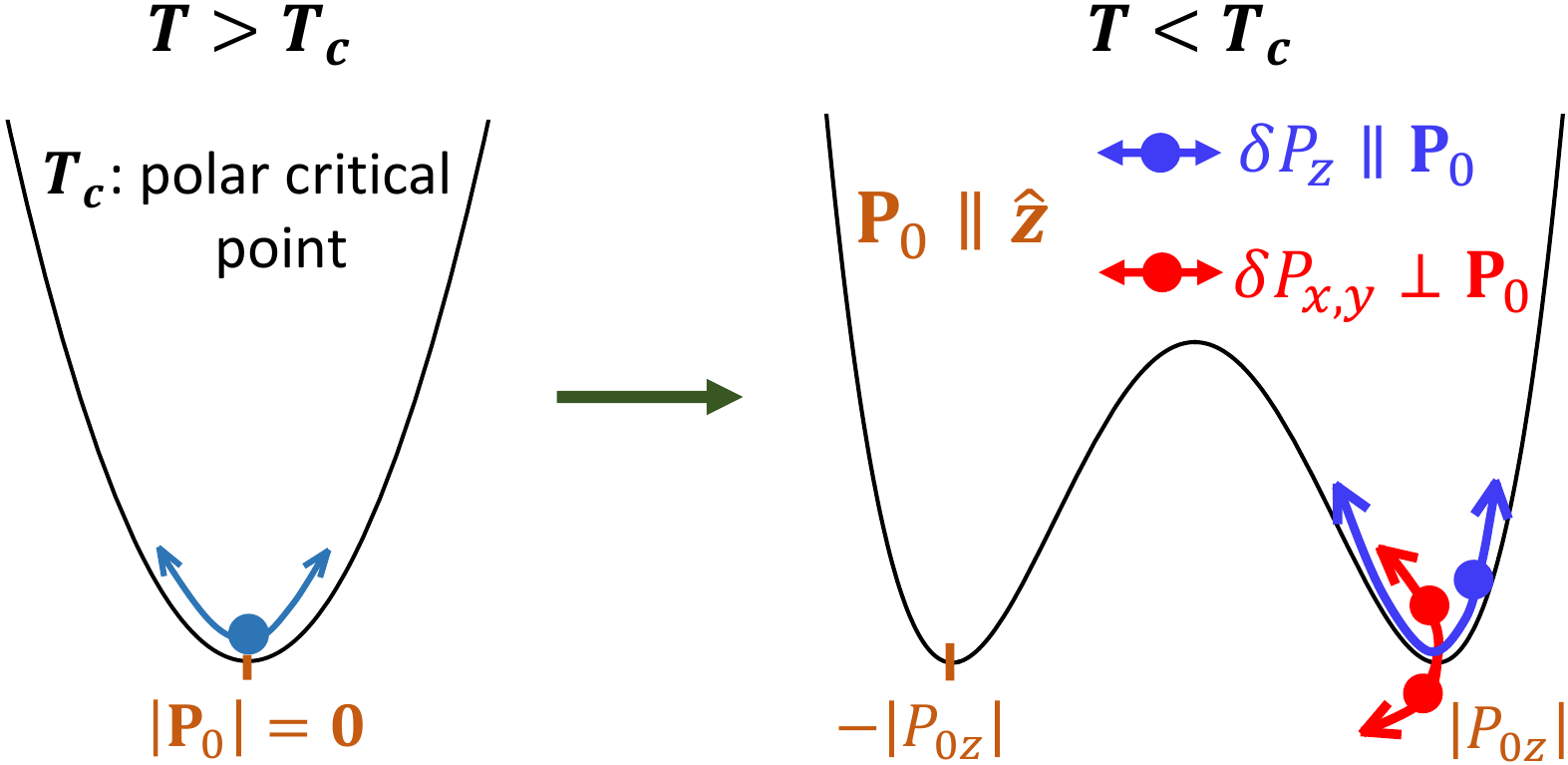}
\caption{\label{fig:GL} Schematics of the phonon excitations above (left) and below (right) the polar $T_c$. 
For $T<T_c$, a macroscopic phonon displacement develops, leading to spontaneous inversion symmetry breaking. The associated phonon excitations become anisotropic, as represented by fluctuations along (solid blue circles) and perpendicular (solid red circles) to the direction of the local polar order ($\bP_0 \parallel \hat{z}$).}
\end{figure}

Since we consider the polar order aligned with the $z$-axis, a simple minimization of the Lagrangian with respect to $P_z$ (assuming $P_x = P_y = 0$) gives
\beq
\label{p0}
P_{0z} = \pm \sqrt{-\frac{w^2}{g+v}},
\eeq
where $w^2<0$, which corresponds to the two ground states of the system emerging due to spontaneous symmetry breaking in the ordered phase, as schematically shown in Fig.~\ref{fig:GL} (right panel). The Fig.~\ref{fig:GL} (left panel) is the representative of non-polar (paraelectric) phase.

Even though the polar order is oriented along $z$-axis, fluctuations in the polar order are present in all directions with
\beq
\label{dPs}
P_x = \delta P_x, \quad\quad P_y = \delta P_y \quad {\rm and} \quad P_z = P_{0z} + \delta P_z,
\eeq
where $P_{0z}$ is given in Eq.~\er{p0}. When we subsitute \er{p0} and \er{dPs} into \er{fe}, we get the imaginary time Lagrangian up to second order in the fluctuations as
\beq
\label{fe1}
\delta{\mathcal L}_{ph}  \approx
\frac{2\pi}{\Omega_0^2} \left(\frac{1}{2} \sum_i \delta \dot{P}_i^2 - w^2 \delta P_z^2 + \frac{v w^2}{2(g+v)} (\delta P_x^2 + \delta P_y^2)
\right),
\eeq
where a constant shift has been absorbed.
We note that the transverse fluctuations have acquired a mass that is proportional to $v$, and thus is physically due to the cubic anisotropy breaking the O(3) rotational symmetry.
Since the coefficients of $\delta P_z^2$ and $\delta P_{x, y}^2$ in \er{fe1} have different
signs, stability of the ordered phase requires that $v$ and $g$ must have different signs: $g>0$, $v<0$ with $g>|v|$.

We can now express the bare phonon propagator in the polar phase (recovering $\bq$-dependence as well for generality) as  
\beq
\label{phgr}
\begin{split}
\mathcal{D}_{\alpha\beta}^0 (\bq, \omega_m) = -\frac{\Omega_0^2}{2\pi} \Bigg( \frac{\delta_{\alpha z}}{\omega_m^2 + 2\omega_\bq^2} &+ \frac{\delta_{\alpha, (x,y)}}{\omega_m^2 + \frac{|v| \omega_\bq^2}{g-|v|}} \Bigg) \\ & \times e_\alpha(\bq) e_\beta(\bq),
\end{split}
\eeq
where $\omega_m$ is the Matsubara frequency, $e_\alpha(\bq) e_\beta(\bq) = \delta_{\alpha\beta}$ is the polarization factor and $\omega_\bq^2=-w^2>0$. 
We next assume $q=0$ for the soft polar modes, and also consider only large carrier density so that
the LO-TO splitting can be neglected \cite{kumar_polar}. Given these assumptions,
the expression for the bare phonon propagator, \er{phgr}, simplifies to
\beq
\label{phgr1}
\mathcal{D}_{\alpha\beta}^0 (\omega_m) = -\delta_{\alpha\beta} \frac{\Omega_0^2}{2\pi} \Bigg( \frac{\delta_{\alpha z}}{\omega_m^2 + \omega_\parallel^2} + \frac{\delta_{\alpha, (x,y)}}{\omega_m^2 + \omega_\perp^2} \Bigg),
\eeq
where $\omega_\parallel = \sqrt{2}\omega_0$ and $\omega_\perp = \omega_0 \sqrt{|v|/(g-|v|)}$, with $\omega_0 \equiv \omega_\bq (q=0)$ as the polar phonon frequency.

\subsection{Spin Susceptibility, ESR and EDSR} \label{RPA}
In this subsection we describe the framework used to calculate the spin-susceptibility; this then allows us to identify and characterize phonon-induced
electronic collective modes and their observable signatures
in ESR experiments. 
More specifically, we demonstrate that the interaction correction to the spin-susceptibility depends on the spin - spin-current correlation. It will be explicitly shown in Sec.~\ref{ESR} that the spin - spin-current correlation is proportional to the static Rasbha parameter $\alpha = \lambda |\bf{P}_0|$.

\begin{figure}
\centering
\includegraphics[scale=0.42]{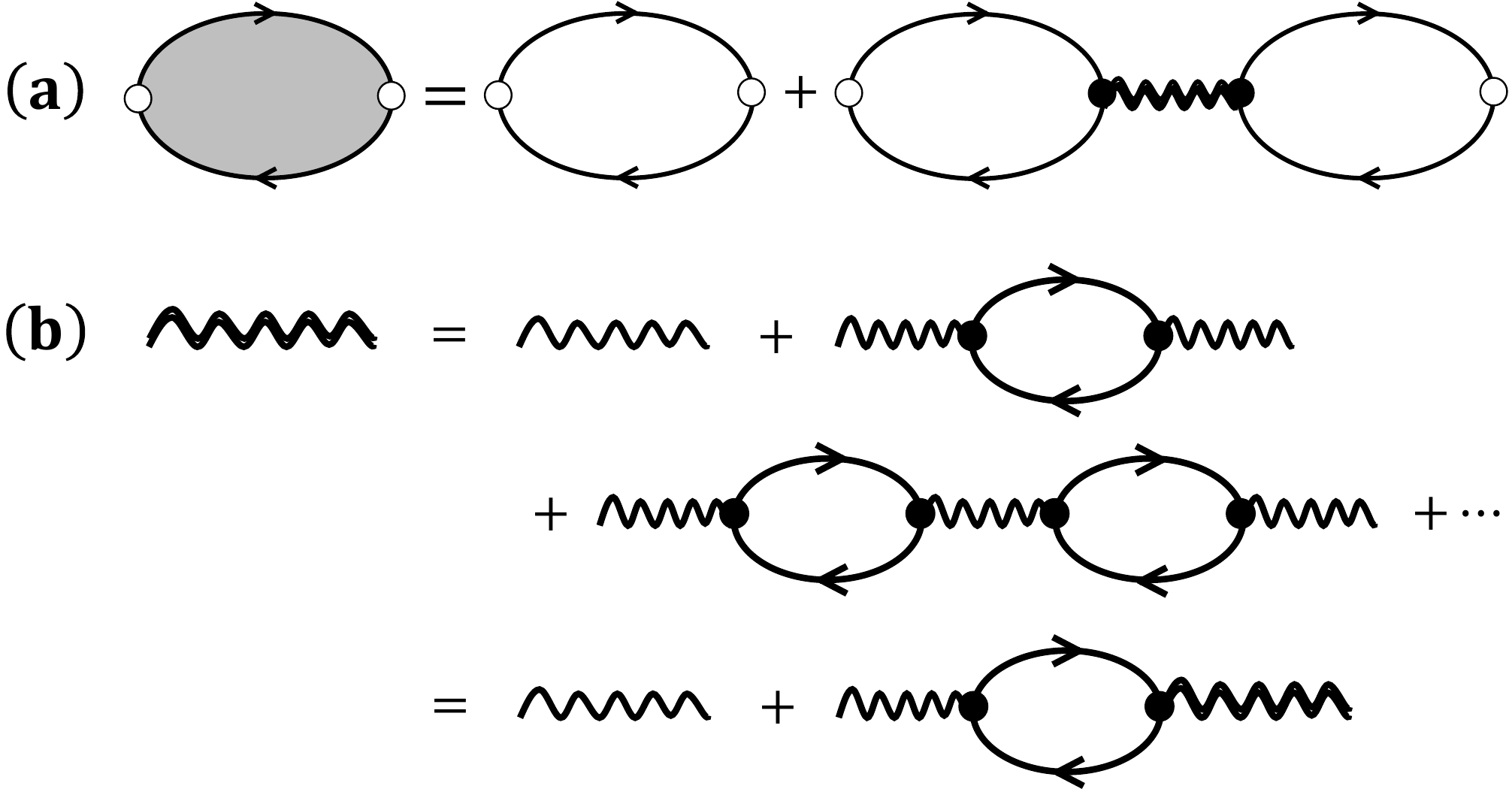}
\caption{\label{fig:RPA} (a) Chain-like RPA sum of $\chi_{ij}$. The open circle is spin ($\bs$) while the solid circle is a spin-current ($\bk \times \bs$) vertex. The double wavy line is the renormalized phonon propagator which enters with spin-current vertex indicating the Rashba type electron-phonon coupling \er{coupling}. (b) Summation scheme for phonon propagator. The single wavy line is the bare phonon propagator \er{phgr1}. The renormalized phonon propagator is obtained by inserting the spin-current - spin-current correlation bubble (indicative of electron-phonon interaction) at successive orders in a chain-like fashion.}
\end{figure}

Ideally, an ESR experiment probes the response of the system to an ac magnetic field that couples to the spins of the electrons only. It is thus expected to be proportional to the spin susceptibility.
However, in a real experiment, the sample is probed by electromagnetic waves having both electric and magnetic field components. Interestingly, in spin-orbit coupled systems, the electric field also couples to the spin response. This is known as electric-dipole spin resonance (EDSR), i.e. the contribution of the spin response to the optical conductivity \cite{rashbaefros2003, efros2006, duckheim2006}. As we show below, the ESR and EDSR responses are related (see Eq.~\er{rel}), so the resonance features in their corresponding response functions occur at same frequencies.
Note however, that typically one expects EDSR to dominate when they overlap in frequency. Indeed, the ratio of ESR to EDSR terms typically is controlled by a small parameter, $(l_C/l_F)^2 \sim 10^{-9} - 10^{-8}$, where $l_F$ is the Fermi wavelength and $l_C = \hbar/mc$ is the Compton length \cite{shekhter2005, maiti2016}.

In subsequent sections we do not discuss the ESR and the EDSR separately, but rather focus on the ESR response (spin-susceptibility) only which is related to that of the EDSR.  
The resonance feature in ESR corresponds to a pole in the imaginary part of the spin-susceptibility, $\chi''_{ij}(\Omega)$, whereas that in EDSR corresponds to a pole in the real part of optical conductivity, $\sigma'_{ij}(\Omega)$. The real part of optical conductivity is related to the imaginary part of current-current correlation ($\mathcal{K}''_{ij}$) as $\sigma'_{ij}(\Omega) = i \mathcal{K}''_{ij}/\Omega$ \cite{mahan:book}. The current operator is related to the velocity operator which is obtained from $\hat{\vec{v}} = i [ \hat{H}_\text{single} + \hat{H}_\text{el-ph}, \hat{\vec{x}} ]$ \cite{maiti2016}, where $\hat{H}_\text{single}$ and $\hat{H}_\text{el-ph}$ are given by Eqs.~\er{single} and \er{coupling}, respectively, with $\bP_\bq$ in the latter is replaced by $\delta\bP_\bq$.

In our model, both the single-particle Hamiltonian (this includes static Rashba) and the interaction Hamiltonian (dynamical Rashba) do not commute with the position operator ($\hat{\vec{x}}$). This indicates that the velocity operator (spin part) will have contributions from both static and dynamical Rashba terms. However, the contribution of the dynamical Rashba term can be ignored compared to the static term for a range of the GL parameters $g$ and $v$. According to Eqs.~\er{stat_rash} and \er{p0}, the coefficient of static Rashba term ($\alpha$) is proportional to $P_{0z} \propto 1/\sqrt{g-|v|}$. If we assume $(g, |v|) \to (ag, a|v|)$, where $a>0$ is some small parameter, then the coefficient of static Rashba term transforms as $\alpha \to \alpha/\sqrt{a}$. Knowing that the dynamical Rashba term eventually leads to phonon propagator, with phonon frequencies as $\omega_\parallel = \sqrt{2}\omega_0$ and $\omega_\perp = \omega_0 \sqrt{|v|/(g-|v|)}$ \er{phgr1}, the similar transformation as above leaves the dynamical term invariant. This provides a clear limit where the static Rashba term would be dominant over the dynamical Rashba term. In what follows, we focus on this case, leaving the study of the dynamical Rashba contribution to the current vertex to future work.

The velocity operator, omitting the dynamical Rashba term, is given by:
\beq
\label{vel}
\hat{\vec{v}} \approx \left( \frac{k_x}{m} \hat{\sigma}_0 + \alpha \hat{\sigma}_y, \frac{k_y}{m} \hat{\sigma}_0 - \alpha \hat{\sigma}_x, \frac{k_z}{m} \hat{\sigma}_0 \right).
\eeq
With the form of velocity operator as above \er{vel}, we find that in the frequency regime relevant for spin resonance features, $\chi''_{ij}(\Omega)$ and $\sigma'_{ij}(\Omega)$ are related by
\beq
\label{rel}
(\sigma'_B)_{11(22)} = \frac{e^2}{(g \mu_B)^2 \Omega} \alpha^2 \chi''_{22(11)},
\eeq
where $\sigma'_B$ is the conductivity component arising solely from magnetic fields (Zeeman and effective magnetic field due to static Rashba SOC). We note that the Drude contribution to the optical conductivity ($\sigma_0'$) is dominant mainly at low frequencies, so its effect at finite frequency regimes where spin resonance features are significant would be weak for a good quality material system. Furthermore, we also observe that Eq.~\er{rel} holds only for in-plane components of $\chi''$ and $\sigma_B'$. 
The reason is that the $z$-component of velocity operator \er{vel} does not have spin character. Note that $z$ corresponds to the polar axis. 
In our model, therefore, $\sigma'_{33}$ (optical conductivity for polarization along the static polar axis) probes only the Drude contribution.

The Eq.~\er{rel} entails an important information which suggests that poles in planar components of the spin-susceptibility also show up in those of the optical conductivity. 
In the following, we will focus our discussion only on spin-susceptibility while referring to Eq.~\er{rel} for the EDSR scenario acknowledging that the absorption rate is determined by $(\sigma'_B)_{11(22)}$ to very high accuracy.

In order to calculate the spin susceptibility within a weak-coupling approximation, we perform a random-phase approximation (RPA) summation of diagrams as illustrated in Fig.~\ref{fig:RPA}. 
The spin susceptibility is determined from a summation of polarization bubbles (Fig.~\ref{fig:RPA}(a)),
and includes spin - spin-current and spin-current - spin bubbles; they are connected by a renormalized phonon propagator, determined by a Dyson equation (Fig.~\ref{fig:RPA}(b)).

Collective modes of an interacting system in the spin sector appear as
poles of the spin-susceptibility. Within the RPA framework of this problem
(cf. Fig.~\ref{fig:RPA}),
the spin susceptibility is expressed as  
\beq
\label{def:susc}
\begin{split}
\chi_{ij} (Q) = &- \frac{(g\mu_B)^2}{4} \bigg[ \Pi_{ij}^{ss}(Q) + \sum_{\alpha\beta} \Pi_{i\alpha}^{ssc}(Q) \mathcal{D}_{\alpha\beta} (Q) \Pi_{\beta j}^{scs}(Q) \bigg],
\end{split}
\eeq
where
\beq
\label{ss}
\Pi_{ij}^{ss}(Q) = \int_K \text{Tr} \big[ \hat{\sigma}_i \hat{G}^0(K-Q/2) \hat{\sigma}_j \hat{G}^0(K+Q/2) \big],
\eeq

\beq
\label{ssc}
\begin{split}
\Pi_{ij}^{ssc}(Q) = \lambda \int_K \text{Tr} \big[ \hat{\sigma}_i \hat{G}^0(K-Q/2) (\bk \times \hat{\bs})_j \hat{G}^0(K+Q/2) \big],
\end{split}
\eeq

\beq
\label{scs}
\begin{split}
\Pi_{ij}^{scs}(Q) = \lambda \int_K \text{Tr} \big[ (\bk \times \hat{\bs})_i \hat{G}^0(K-Q/2) \hat{\sigma}_j \hat{G}^0(K+Q/2) \big],
\end{split}
\eeq
are the bare spin-spin, spin - spin-current and spin-current - spin correlation functions respectively, and
\beq
\label{full_ph}
\hat{\mathcal{D}}(Q) = \big[ [\hat{\mathcal{D}}^0(Q)]^{-1} - \hat{\Pi}^{scsc}(Q) \big]^{-1},
\eeq
is the full RPA phonon propagator ((cf. Fig.~\ref{fig:RPA} (b)). In Eq.~\er{full_ph}, $\mathcal{D}_{ij}^0$ is
the bare phonon propagator \er{phgr1} and the self-energy correction, the spin-current - spin-current correlation function, included is
\beq
\label{scsc}
\begin{split}
\Pi_{ij}^{scsc}(Q) = \lambda^2 \int_K \text{Tr} \big[ & (\bk \times \hat{\bs})_i \hat{G}^0(K-Q/2) \\
\times & (\bk \times \hat{\bs})_j \hat{G}^0(K+Q/2) \big].
\end{split}
\eeq
In Equations \er{ss}-\er{scsc}, $Q = (i\Omega_n, \bq)$, $K = (i\omega_m, \bk)$, $\int_K \equiv T\sum_{\omega_m} \int d^3k/(2\pi)^3$, $\epsilon_m$ is the Matsubara frequency, and
\beq
\label{gr}
\begin{split}
\hat{G}^0(i\epsilon_m, & \bk) = \sum_r \frac{\ket{r} \bra{r}}{i\epsilon_m + \mu - \ve_\bk^r} = \sum_r \hat{D}_r(\bk) g_r(i\epsilon_m, \bk), \\
\hat{D}_r(\bk) &= \frac{1}{2} \bigg[ \hat{\sigma}_0 + r \frac{\Delta_Z}{\Delta_\bk} \hat{\sigma}_z + r \frac{2\alpha}{\Delta_\bk} (\hat{\sigma}_y k_x - \hat{\sigma}_x k_y) \bigg], \\
g_r(i\epsilon_m, & \bk) = \frac{1}{i\epsilon_m + \mu - \ve_\bk^r}
\end{split}
\eeq
with $\mu$ as the chemical potential, $\alpha = \lambda |P_{0z}|$ and $g_r(i\epsilon_m, \bk)$ \er{gr} identified as a chiral Green's function; the eigenvalues ($\ve_\bk^r$), eigenstates ($\ket{r}$) and $\Delta_\bk$ appearing  in \er{gr} are given in Eq.~\er{es}.

The approximation in Fig.~\ref{fig:RPA}(a) captures only a subset of diagrams for the spin response. However, the
diagrams we select contain singularities at the frequencies of the eigenmodes, which are determined accurately to order $\lambda^2$ by those in Fig.~\ref{fig:RPA}(b). The diagrams neglected in Fig.~\ref{fig:RPA}(a) will not contribute any additional singularities or shift the eigenmode frequencies at weak coupling, but can alter the spectral weights associated with them. Therefore, we use the approximation in Fig.~\ref{fig:RPA}(a) to assess the qualitative form of the spin response spectrum, while acknowledging that a quantitative description of its intensity requires further analysis which is beyond the scope of this work.

We next discuss collective modes of the spin-susceptibility \er{def:susc}; most of its poles are associated with the spin-flip particle-hole continuum. 
The only contribution that contains
interaction effects is the renormalized phonon propagator,
$\mathcal{D}_{ij}$, as defined in Eq.~\er{full_ph}.
Poles in $\mathcal{D}_{ij}$ result in poles in $\chi_{ij}$ if
the spin - spin-current bubble, $\Pi_{ij}^{ssc}$, is finite; only
then are these collective modes excited. We emphasize
that coupling between the spin and the spin-current degrees of
freedom is only possible in the ordered phase since the latter
but not the former breaks inversion symmetry.

Calculation of the phonon self-energy $\Pi_{ij}^{scsc}$ \er{scsc} is thus central towards the identification of spin-active collective modes.
Using \er{scsc} with $\mathcal{D}_{ij}^0$ \er{phgr1}, one can determine the renormalized phonon propagator $\mathcal{D}_{ij}$ \er{full_ph} and then, via $\chi_{ij}$ \er{def:susc}, the spin-active collective modes. 
Throughout this work we will assume a $q=0$ soft mode, and no distinction between the LO and TO modes in the ordered phase.

Let us now simplify whenever possible the expressions needed to characterize these
interaction-induced collective
mode.  The phonon self-energy $\Pi_{ij}^{scsc}$ can be expressed in the following compact form
\beq
\label{compact}
\Pi_{ij}^{scsc}(i\Omega_n) = \lambda^2 \int_K \sum_{r \bar{r}} f_{ij}^{r\bar{r}, scsc} g_r (i(\omega_m + \Omega_n), \bk) g_{\bar{r}} (i\omega_m, \bk),
\eeq
where 
$g_r(i\epsilon_m, \bk)$ is given by Eq.~\er{gr} and $f_{ij}^{r\bar{r}, scsc}$ is the coherence factor of the spin-current - spin-current correlation function.
For brevity, we have omitted the arguments of $f_{ij}^{r\bar{r}, scsc}$ which are mainly polar and azimuthal angles. The explicit form of $f_{ij}^{r\bar{r}, scsc}$ is presented in Eq.~\er{coh_scsc} of Appendix~\ref{appen:coh}.

Next we exploit the symmetries of our problem to simplify the different matrix
elements of $\chi_{ij}$. Since the applied magnetic field and the polar order
are both aligned with $z$-axis, 
the system preserves $C_2$ rotation about the $z$-axis, which transforms $x \to -x$ and $y \to -y$ simultaneously.
This leaves both $\hat{\sigma}_z$ and
$(\bk \times \hat{\bs})_z$ invariant, while $\hat{\sigma}_{x, y}$ and $(\bk \times \hat{\bs})_{x, y}$ change sign. This has consequences for
$\Pi_{ij}^{scsc}$, $\Pi_{ij}^{ss}$, $\Pi_{ij}^{ssc}$ and $\Pi_{ij}^{scs}$.
For example,  because $\Pi_{ij}^{scsc}$
involves both $(\bk \times \hat{\bs})_i$
and $(\bk \times \hat{\bs})_j$, only $\Pi_{12}^{scsc}$ and the diagonal elements
$\Pi_{ii}^{scsc}$ are nonvanishing under this symmetry.
Finally, the system's rotational symmetry in the x-y plane further leads
to $\Pi_{11}^{scsc} = \Pi_{22}^{scsc}$.

We can therefore express $\mathcal{D}_{ij}$ \er{full_ph} in a simplified form based on these symmetry considerations:
\beq
\label{com:phgr}
\begin{split}
\mathcal{D}_{11} &= \frac{\mathcal{D}_{11}^0 - \Pi_{11}^{scsc} (\mathcal{D}_{11}^0)^2}{\big( 1 - \Pi_{11}^{scsc} \mathcal{D}_{11}^0 \big)^2 + (\Pi_{12}^{scsc})^2 (\mathcal{D}_{11}^0)^2}, \\
\mathcal{D}_{22} &= \mathcal{D}_{11}, \\
\mathcal{D}_{33} &= \frac{\mathcal{D}_{33}^0}{1 - \Pi_{33}^{scsc} \mathcal{D}_{33}^0}, \\
\mathcal{D}_{12} &= \frac{\Pi_{12}^{scsc} (\mathcal{D}_{11}^0)^2}{\big( 1 - \Pi_{11}^{scsc} \mathcal{D}_{11}^0 \big)^2 + (\Pi_{12}^{scsc})^2 (\mathcal{D}_{11}^0)^2}, \\
\mathcal{D}_{21} &= -\mathcal{D}_{12}.
\end{split}
\eeq
Since the collective modes are poles of $\hat{\mathcal{D}}$, they can be obtained by solving $\text{Det}[\hat{\mathcal{D}}^{-1}]=0$ for frequencies. According to Eq.~\er{com:phgr}, this condition becomes
\beq
\label{det}
\begin{split}
\text{Det}[\hat{\mathcal{D}}^{-1}] &= \Big[ (\Pi_{12}^{scsc})^2 + \big( \Pi_{11}^{scsc} - (\mathcal{D}_{11}^0)^{-1} \big)^2 \Big] \times \\
& \hspace{2cm} \times \big[ -\Pi_{33}^{scsc} + (\mathcal{D}_{33}^0)^{-1} \big]. \\
&= 0 
\end{split}
\eeq

We now turn to a discussion of the spin-susceptibility, which also 
has contributions from 
$\Pi_{ij}^{ss}$, $\Pi_{ij}^{ssc}$ and $\Pi_{ij}^{scs}$
according to Eq.~\er{def:susc}. These bubbles
can be written in the same compact form as $\Pi_{ij}^{scsc}$ \er{compact},
where the corresponding coherence factors, $f_{ij}^{r\bar{r}, ss}$, $f_{ij}^{r\bar{r}, ssc}$ and $f_{ij}^{r\bar{r}, scs}$, are given in Eqs.~\er{coh_ss}, \er{coh_ssc} and \er{coh_scs} of Appendix~\ref{appen:coh}.
Again exploiting the symmetry associated with the $z$-axis alignment of
the polar order and the magnetic field, the only non-zero elements
of the spin-susceptibility \er{def:susc}
are $\chi_{12}$, $\chi_{21}$ and its diagonal components $\chi_{ii}$
so that we can now write 
\beq
\label{spin-susc}
\begin{split}
\chi_{11} = -\big[ \Pi_{11}^{ss} &+ \Pi_{11}^{ssc} \mathcal{D}_{11} \Pi_{11}^{scs} + \Pi_{11}^{ssc} \mathcal{D}_{12} \Pi_{21}^{scs} \\
&+ \Pi_{12}^{ssc} \mathcal{D}_{21} \Pi_{11}^{scs} + \Pi_{12}^{ssc} \mathcal{D}_{22} \Pi_{21}^{scs} \big], \\
\chi_{22} = -\big[ \Pi_{22}^{ss} &+ \Pi_{21}^{ssc} \mathcal{D}_{11} \Pi_{12}^{scs} + \Pi_{21}^{ssc} \mathcal{D}_{12} \Pi_{22}^{scs} \\
&+ \Pi_{22}^{ssc} \mathcal{D}_{21} \Pi_{12}^{scs} + \Pi_{22}^{ssc} \mathcal{D}_{22} \Pi_{22}^{scs} \big], \\
\chi_{33} = -\big[ \Pi_{33}^{ss} &+ \Pi_{33}^{ssc} \mathcal{D}_{33} \Pi_{33}^{scs} \big], \\
\chi_{12} = -\big[ \Pi_{12}^{ss} &+ \Pi_{11}^{ssc} \mathcal{D}_{11} \Pi_{12}^{scs} + \Pi_{11}^{ssc} \mathcal{D}_{12} \Pi_{22}^{scs} \\
&+ \Pi_{12}^{ssc} \mathcal{D}_{21} \Pi_{12}^{scs} + \Pi_{12}^{ssc} \mathcal{D}_{22} \Pi_{22}^{scs} \big], \\
\chi_{21} = -\big[ \Pi_{21}^{ss} &+ \Pi_{21}^{ssc} \mathcal{D}_{11} \Pi_{11}^{scs} + \Pi_{21}^{ssc} \mathcal{D}_{12} \Pi_{21}^{scs} \\
&+ \Pi_{22}^{ssc} \mathcal{D}_{21} \Pi_{11}^{scs} + \Pi_{22}^{ssc} \mathcal{D}_{22} \Pi_{21}^{scs} \big],
\end{split}
\eeq
where $\chi_{ij} \equiv \chi_{ij}/(g^2\mu_B^2/4)$. 

The presence of rotational symmetry allows for even further simplification of Eq.~\er{spin-susc} that results in the expressions 
\beq
\label{susc1}
\begin{split}
\chi_{11} &= -\big[ \Pi_{11}^{ss} + \big( (\Pi_{11}^{ssc})^2 - (\Pi_{12}^{ssc})^2 \big) \mathcal{D}_{11} - 2\Pi_{11}^{ssc} \Pi_{12}^{ssc} \mathcal{D}_{12} \big], \\
\chi_{22} &= \chi_{11}, \\
\chi_{33} &= -\big[ \Pi_{33}^{ss} + (\Pi_{33}^{ssc})^2 \mathcal{D}_{33} \big], \\
\chi_{12} &= -\big[ \Pi_{12}^{ss} + \big( (\Pi_{11}^{ssc})^2 - (\Pi_{12}^{ssc})^2 \big) \mathcal{D}_{12} + 2\Pi_{11}^{ssc} \Pi_{12}^{ssc} \mathcal{D}_{11} \big], \\
\chi_{21} &= -\chi_{12},
\end{split}
\eeq
where the $\mathcal{D}_{ij}$ are given in Eq.~\er{com:phgr}. 

In order to identify and characterize interaction-induced collective modes of the polar metal, we will use the framework outlined here and will determine explicit forms of the expressions in \er{com:phgr} and \er{susc1}. The modes induced by the electron-phonon interaction will appear as poles in $\hat{\mathcal{D}}$; they will only be accessible by ESR in the (ordered) polar metal where $\Pi_{ij}^{ssc}$ is finite. 

\section{Interaction-Induced Bound States}
\label{coll_mod}
In this section we explicitly determine the electronic collective
modes due to spin-orbit assisted electron-phonon coupling in an applied
magnetic field. These modes correspond to the poles of the phonon propagator renormalized by the electron-phonon interactions, Eq.~\er{coupling}. 
The resulting phonon self-energy is related to the spin-current - spin-current
polarization tensor $\Pi_{ij}^{scsc}$. 
Divergences in the latter indicate enhancement of interaction effects, and we first study them both at zero and at finite magnetic fields. 
Next we analyze the renormalized phonon propagator and demonstrate the emergence of new collective bound states near the edges of the particle-hole continuum even at weak coupling.
Furthermore we show that an applied field parallel to the polar axis tunes the number of these collective modes and their energies. We emphasize that the origin of theses bound states is the
Rashba spin-orbit type electron-phonon interaction.

\subsection{Spin-current - spin-current polarization tensor}
\label{spin-current_tensor}
In order to calculate collective modes, we require forms of $\mathcal{D}_{ij}$ \er{com:phgr} which depend on self-energy corrections $\Pi_{ij}^{scsc}$. 
Assuming $q=0$ and Fermi energy to be the largest energy scale in the system, the non-zero components of $\hat{\Pi}^{scsc}$ can be calculated:
\beq
\label{Pi_scsc}
\begin{split}
\Pi_{11}^{scsc} (\Omega_n) &= - A \Bigg\{ 2 - \frac{4\Delta_Z^2}{\Delta_R^2} + \frac{(-\Omega_n^2 + \Delta_Z^2)}{\Delta_R \sqrt{\Omega_n^2 + \Delta_R^2 + \Delta_Z^2}} L(\Omega_n) \\
&\hspace{1.5cm} + \frac{2\Delta_Z^2 (\Omega_n^2 + \Delta_Z^2)}{\Delta_R^3 \sqrt{\Omega_n^2 + \Delta_R^2 + \Delta_Z^2}} L(\Omega_n) \Bigg\}, \\
\Pi_{22}^{scsc} (\Omega_n) &= \Pi_{11}^{scsc} (\Omega_n), \\
\Pi_{33}^{scsc} (\Omega_n) &= - A \frac{\Delta_Z^2}{\Delta_R^2} \Bigg\{ 4 - \frac{2(\Omega_n^2 + \Delta_Z^2)}{\Delta_R \sqrt{\Omega_n^2 + \Delta_R^2 + \Delta_Z^2}} L(\Omega_n) \Bigg\}, \\
\Pi_{12}^{scsc} (\Omega_n) &= - A \frac{4\Omega_n \Delta_Z}{\Delta_R^2} \Bigg\{ 1 - \frac{\sqrt{\Omega_n^2 + \Delta_R^2 + \Delta_Z^2}}{2\Delta_R} L(\Omega_n) \Bigg\}, \\
\Pi_{21}^{scsc} (\Omega_n) &= -\Pi_{12}^{scsc} (\Omega_n),
\end{split}
\eeq
where
\beq
\label{L}
L(\Omega_n) = \text{log} \Bigg[ \frac{\sqrt{\Omega_n^2 + \Delta_R^2 + \Delta_Z^2} + \Delta_R}{\sqrt{\Omega_n^2 + \Delta_R^2 + \Delta_Z^2} - \Delta_R} \Bigg],\,\, A \equiv \frac{m k_F^3 \lambda^2}{4\pi^2},
\eeq
and $\Delta_R \equiv 2\alpha k_F$, with $k_F$ as the Fermi momentum and $N_F=mk_F/\pi^2$ as the total density of states in 3D, is the energy scale set by static Rashba SOC. The effective coupling constant $A$, in its current form, is dimensionless.
The emergence of the static polar order parameter \er{p0} in the ordered phase allows us to write the static Rashba parameter ($\alpha$) in Eq.~\er{stat_rash}, according to the relation $w^2 = -\omega_0^2$, and Eqs.~\er{p0} and \er{fe1} (regarding the sign of $v$), as
\beq
\label{alpha}
\alpha = \lambda |P_{0z}| = \lambda \sqrt{\frac{\omega_0^2}{g-|v|}},
\eeq
where $g>|v|$. Following Eqs.~\er{alpha} and \er{phgr1}, the Rashba energy can be then written as
\beq
\label{rash}
\Delta_R \equiv 2\alpha k_F = 2\lambda k_F \sqrt{\frac{\omega_\parallel^2}{2(g-|v|)}} = 2 \omega_\parallel \sqrt{\frac{2A}{(g-|v|)N_F}},
\eeq
where $A$ is the effective coupling constant as defined in Eq.~\er{L}.
From rotational symmetry, $\Pi_{22}^{scsc} (i\Omega_n) = \Pi_{11}^{scsc} (i\Omega_n)$ and $\Pi_{21}^{scsc} (i\Omega_n) = -\Pi_{12}^{scsc} (i\Omega_n)$. The technical details of obtaining Eq.~\er{Pi_scsc} is delegated to Appendix~\ref{appen:bubble}, while the main results and their qualitative features are discussed in the main text. 

The log[...] in $\Pi_{ij}^{scsc}$ \er{Pi_scsc} appears mainly because of the effective two-dimensionality of the $\bk$-integral which appears in its definition \er{compact}. To demonstrate this, let's first assume the coherence factor appearing in the integrand of \er{compact} to be one. We are now left with two chiral Green's functions, $g_r(i\epsilon_n, \bk)$ \er{gr}, for frequency summation and then $\bk$-integral. Upon frequency summation, Eq.~\er{compact} can be written as
\beq
\label{demo1}
\Pi_{ij}^{scsc} \sim \sum_{r, \bar{r}} \int \frac{d^3 k}{(2\pi)^3} \frac{n_F(\ve_\bk^{\bar{r}} - \mu) - n_F(\ve_\bk^r - \mu)}{i\Omega_n - \ve_\bk^r + \ve_\bk^{\bar{r}}}.
\eeq
Now dividing $\bk$ into its parallel and perpendicular components, with the assumption that the perpendicular component is projected onto Fermi surface, we simplify the  form of Eq.~\er{demo1} to be 
\beq
\label{demo2}
\begin{split}
\Pi_{ij}^{scsc} \sim& \sum_{r, \bar{r}} \int \frac{dk_\perp}{2\pi} \int \frac{d^2k_\parallel}{(2\pi)^2} \frac{(\bar{r} - r) \sqrt{\Delta_Z^2 + 4\alpha^2 k_\parallel^2}}{i\Omega_n + (\bar{r} - r) \sqrt{\Delta_Z^2 + 4\alpha^2 k_\parallel^2}} \times \\
&\hspace{4cm} \times \delta \big( k_\perp^2/2m - \mu \big).
\end{split}
\eeq
The $k_\perp$ above is immediately projected onto $k_F$, as ensured by the $\delta$-function. Knowing that the $k_\parallel$-integral is also truncated by $k_F$ in the upper limit, one can easily see that the $k_\parallel$ integration in Eq.~\er{demo2} yields a log[..] (using a power counting argument for $k$), which is consistent with the results in Eqs.~\er{Pi_scsc} and \er{L}. Indeed, this comes from the dimensional reduction of the $\bk$-integral from three to two dimensions.

As mentioned earlier in this section, the divergence in $\Pi_{ij}^{scsc}$ enhances the interaction effects. To understand this better, we first discuss the limiting case of $\Delta_Z=0$ in Sec.~\ref{scsc_Z=0}, followed by the finite $\Delta_Z$ case in Sec.~\ref{sec:se_Z}.

\subsubsection{Zero magnetic field}
\label{scsc_Z=0}
Components $\Pi_{12}^{scsc}$ and $\Pi_{33}^{scsc}$ vanish at zero magnetic field, see Eq.~\er{Pi_scsc}. At $\Delta_Z=0$, the system is time-reversal symmetric which ensures $\Pi_{12}^{scsc}=0$. Moreover, $\Pi_{33}^{scsc}=0$ follows from spin-current conservation along the direction of polar order, $(\bk \times \hat{\bs})_z$. Indeed, the single-particle Hamiltonian \er{single} contains static Rashba term \er{stat_rash}, with $\hat{n} \parallel \hat{z}$, which has the same form as the spin-current vertex along $\hat{z}$-direction. This tells us that the spin-current along the direction of the polar order is conserved, and thus $\Pi_{33}^{scsc}=0$. Therefore, only $\Pi_{11}^{scsc}$ and $\Pi_{22}^{scsc}$ components (from rotational symmetry $\Pi_{22}^{scsc} = \Pi_{11}^{scsc}$) survive in this limit which reads
\beq
\label{seZ=0}
\begin{split}
\Pi_{11}^{scsc} & (\Omega_n, \Delta_Z=0) = \Pi_{22}^{scsc} (\Omega_n, \Delta_Z=0) \\
=& - A \Bigg\{ 2 - \frac{\Omega_n^2}{\Delta_R \sqrt{\Omega_n^2 + \Delta_R^2}} \text{log} \Bigg[ \frac{\sqrt{\Omega_n^2 + \Delta_R^2} + \Delta_R}{\sqrt{\Omega_n^2 + \Delta_R^2} - \Delta_R} \Bigg] \Bigg\},
\end{split}
\eeq
where $A$ is an effective dimensionless coupling constant as defined in Eq.~\er{L}, and $\Delta_R$ \er{rash} is the Rashba energy. It corresponds to the spin splitting emergent from static polarization in the polar phase given by Eq.~\er{stat_rash}, where $\alpha$ is given by Eq.~\er{alpha}.

\begin{figure*}
\centering
\includegraphics[scale=0.55]{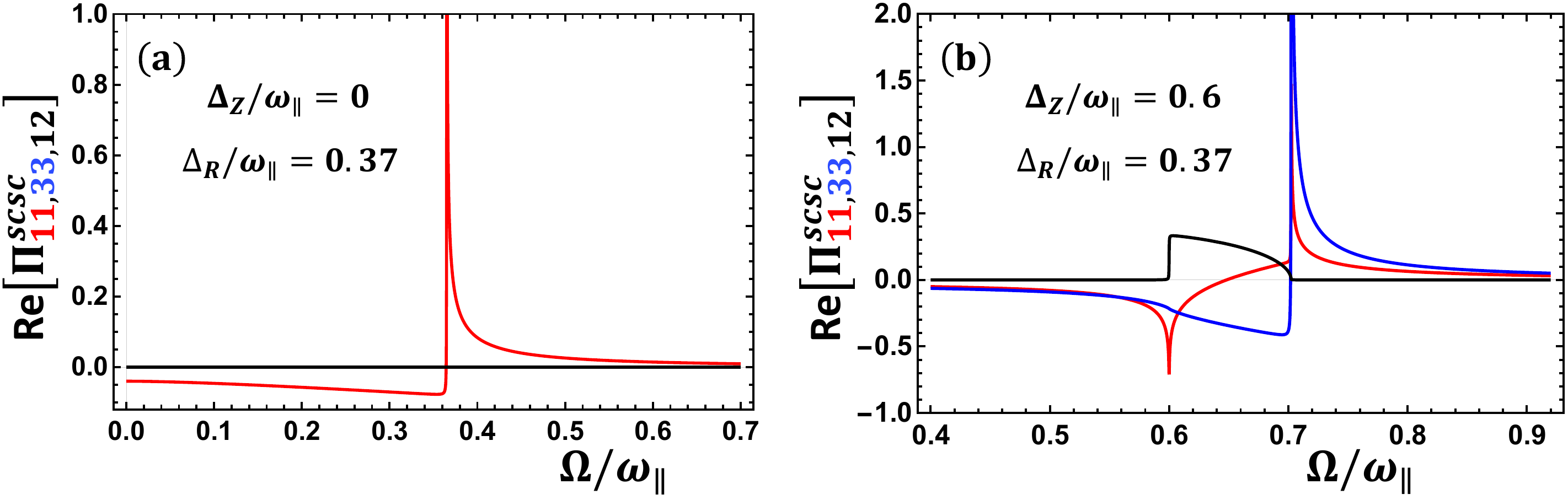}
\caption{\label{fig:pi_scsc} Real part of phonon self-energy \er{full_ph}, or spin-current - spin-current polarization tensor \er{scsc}, at zero (a) and finite (b) Zeeman field. 
The model parameters are taken to be $gN_F=2.4$, $|v|N_F=1.2$ and $A=0.02$, where $N_F=mk_F/\pi^2$ is the total 3D density of states. 
The Rashba energy $\Delta_R$ \er{rash} is determined by $A$ as well as Ginzburg-Landau parameters $g$ and $|v|$. 
Divergences point to strongly enhanced role of interactions around the corresponding frequencies, as is discussed in the text. 
Red, blue and black curves represent $\Pi_{11}^{scsc}$, $\Pi_{33}^{scsc}$ and $\Pi_{12}^{scsc}$, corresponding to $xx$, $zz$ and $xy$ phonon polarization orientations ($yy$ response is identical to $xx$ one). 
(a) At $\Delta_Z = 0$, $\Pi_{33}^{scsc}$ and $\Pi_{12}^{scsc}$ components are zero, as can be seen from Eq.~\er{Pi_scsc}. 
The divergence in Re$\Pi_{11}^{scsc}$ occurs at $\Omega = \Delta_R$, the upper edge of the particle-hole continuum. 
(b) At finite $\Delta_Z$, the continuum is gapped and the divergence at the lower edge, $\Omega = \Delta_Z$, appears in $\Pi_{11}^{scsc}$ and $\Pi_{12}^{scsc}$. 
At the upper edge of the continuum, $\Omega = (\Delta_R^2 + \Delta_Z^2)^{1/2}$, however, the divergence is present only in the diagonal components, $\Pi_{11}^{scsc}$ and $\Pi_{33}^{scsc}$.}
\end{figure*}

As we can see from Eq.~\er{seZ=0}, at $\Delta_Z=0$ $\Pi_{11}^{scsc}$ is square-root divergent at $\Omega=\Delta_R$, where $\Omega$ is the retarded frequency written after analytic continuation $i\Omega_n \to \Omega+i0^+$. 
This divergence is reflected in the real part of $\Pi_{11}^{scsc}$, as shown in Fig.~\ref{fig:pi_scsc}(a). 
The square-root divergence allows for strong interaction effects present already at weak coupling and in particular in the appearance of collective modes which we will discuss in detail in Sec.~\ref{sec:zero_field}.

Another important feature of the spin-current polarization tensor is the emergence of an electronic excitation continuum in the polar phase.
Its presence is revealed at frequencies where the imaginary part of $\Pi_{11}^{scsc}$ is nonzero, or Im$\Pi_{11}^{scsc} \neq 0$. To obtain it, one can either perform analytic continuation of Eq.~\er{seZ=0} or, alternatively, directly calculate Im$\Pi_{11}^{scsc}$ from Eq.~\er{compact} in the limit of $\Delta_Z=0$; it is much easier to do the latter. We obtain Im$\Pi_{11}^{scsc}$ at $\Delta_Z=0$ as
\beq
\label{im11}
\text{Im}\Pi_{11}^{scsc} = - \pi A \frac{\Omega^2}{\Delta_R \sqrt{\Delta_R^2 - \Omega^2}} \Theta(\Omega) \Theta(\Delta_R - \Omega).
\eeq
In the above equation \er{im11}, the constraints imposed by $\Theta$-functions suggest the range of the particle-hole continuum to be $0 < \Omega < \Delta_R$. Moreover, the continuum response is suppressed as $\Omega^2$ at low frequencies, while it is square-root singular at $\Omega = \Delta_R$.

An easier way to understand the location of edges of the continuum is by rewriting eigenvalues \er{es} of the single-particle Hamiltonian \er{single} at $\Delta_Z=0$ and calculating the interband transition frequency:
\beq
\label{trans}
\hbar \Omega = \ve_+ - \ve_- = |2\alpha k \sin\theta|.
\eeq
Assuming Rashba splitting to be small compared to the Fermi energy, it is legitimate to project $k$ onto $k_F$. 
Recognizing $2\alpha k_F$ as $\Delta_R$ \er{rash}, with $\Delta_R>0$, and knowing that $|\sin\theta|$ is bounded between 0 and 1, the minimum and maximum values of $\Omega$ comes out to be $0$ and $\Delta_R$, respectively. These define lower and upper edges of the spin-flip continuum, correspondingly. 

Since the continuum starts right from $\Omega=0$, there is no gap for particle-hole excitations between spin-split subbands. Note that it has been found previously that a weak repulsive inter-electron interaction does not lead to the formation of collective modes in this case \cite{maiti2014}. However, as is discussed below, for electron-phonon interaction and for $\Delta_R\ll E_F$ the appearance of divergence at the upper edge of the continuum $(\Omega = \Delta_R)$ suggests that collective modes (or poles in $\mathcal{D}_{ij}$) may exist at weak coupling above the continuum.

The structure of the matrix bubble ($\hat{\Pi}^{scsc}$) at $\Delta_Z=0$ has interesting consequences: 
the additional vanishing of $\Pi_{33}^{scsc}$ and $\Pi_{12}^{scsc}$ at $\Delta_Z=0$, 
and the divergence feature of the remaining non-zero components, 
$\Pi_{11}^{scsc}$ and $\Pi_{22}^{scsc}$ (with $\Pi_{11}^{scsc} = \Pi_{22}^{scsc}$) \er{seZ=0}, 
already tell us about the qualitative features of collective mode. 
First, it is clear from symmetry (see Sec.~\ref{RPA} for discussion) that the in-plane (\{x\} and \{y\}) and the out-of-plane (\{z\}) sectors are decoupled. 
Moreover, $\Pi_{33}^{scsc}=0$ at $\Delta_Z=0$. 
Therefore, the collective mode in the \{z\}-sector does not exist and we always have a bare phonon polarized along the $z$-direction ($\omega_\parallel$, as defined in Eq.~\er{phgr1}) in this sector. 
The bubbles in the in-plane sector, however, are non-zero with their real part being divergent at $\Omega=\Delta_R$, the upper edge of the continuum. Therefore, we expect collective modes above the continuum in this sector due to interaction between $x$- and $y$-polarized phonons located at $\Omega=\omega_\perp$, with $\omega_\perp$ defined in Eq.~\er{phgr1}, and the electronic continuum. 
Finally, because the time-reversal symmetry is intact at $\Delta_Z=0$, which results in $\Pi_{ij}^{scsc}=0, (i, j) \in \{1, 2\}$, the mode in the in-plane sector is double-degenerate. This is all consistent with the explicit calculation which we will discuss in Sec.~\ref{sec:zero_field}.

\subsubsection{Finite magnetic fields}
\label{sec:se_Z}
Let us now discuss the effects of finite $\Delta_Z$ by analysing Eq.~\er{Pi_scsc}. 
One can notice that unlike for $\Delta_Z=0$ different components of $\Pi_{ij}^{scsc}$ now possess up to two kinds of divergences: the log-divergence at $\Omega=\Delta_Z$ and the square-root divergence at $\Omega=\sqrt{\Delta_R^2+\Delta_Z^2}$. 
While at $\Omega = \Delta_Z$ the log-divergence occurs in components $\Pi_{11}^{scsc}$ and $\Pi_{12}^{scsc}$ (as evident from Eq.~\er{L}), at $\Omega = \sqrt{\Delta_R^2 + \Delta_Z^2}$ the square-root divergence occurs in $\Pi_{11}^{scsc}$ and $\Pi_{33}^{scsc}$.  
We note that the log-divergence in $\Pi_{33}^{scsc}$ (see Eq.~\er{Pi_scsc}) disappears because the log-term \er{L} of this component also accompanies an overall prefactor $(\Omega^2 - \Delta_Z^2)$ (in retarded representation); clearly, this prefactor diminishes the effect of log-term at $\Omega=\Delta_Z$.

All the divergences at finite fields are present in Re$\Pi_{ij}^{scsc}$, which is shown in Fig.~\ref{fig:pi_scsc}(b). 
The Re$\Pi_{11}^{scsc}$ shows both log- and square-root divergences at $\Omega = \Delta_Z$ and $\Omega = \sqrt{\Delta_R^2 + \Delta_Z^2}$, respectively; see Eqs.~\er{Pi_scsc} and \er{L} and also the red curve in Fig.~\ref{fig:pi_scsc}(b). 
The Re$\Pi_{12}^{scsc}$ shows only log-divergence at $\Omega = \Delta_Z$. We note that since $\Pi_{12}^{scsc}$ component is proportional to $\Omega_n$ \er{Pi_scsc}, the log-divergence feature is actually reflected in the imaginary part instead of the real part which shows step-like feature as presented by black curve in Fig.~\ref{fig:pi_scsc}(b). 
Finally, the Re$\Pi_{33}^{scsc}$ shows only the square-root divergence at $\Omega = \sqrt{\Delta_R^2 + \Delta_Z^2}$  as shown by blue curve in Fig.~\ref{fig:pi_scsc}(b): the seeming log-divergence at $\Omega=\Delta_Z$ in Eq.~\er{Pi_scsc} disappears because of the overall prefactor ($\Omega^2-\Delta_Z^2$) in front of the log term.

The divergences in the real part of $\Pi_{ij}^{scsc}$ allow for strong interaction effects at weak coupling strength. Let us now discuss the imaginary part, which is non-zero inside the particle-hole continuum. As discussed in the context of $\Delta_Z=0$ in Sec.~\ref{scsc_Z=0}, the continuum is defined by the region where $\text{Im}\Pi_{ij}^{scsc} \neq 0$. At finite $\Delta_Z$, the $\text{Im}\Pi_{ij}^{scsc}$ can be calculated:
\beq
\label{ImPi_Z}
\begin{split}
\text{Im}\Pi_{11}^{scsc} &= -\pi A \frac{\Delta_R^2(\Omega^2+\Delta_Z^2) - 2\Delta_Z^2(\Omega^2-\Delta_Z^2)}{\Delta_R^3\sqrt{\Delta_R^2+\Delta_Z^2-\Omega^2}} \\
&\hspace{0.8cm} \times \Theta(\Omega-\Delta_Z) \Theta \Big( \sqrt{\Delta_R^2+\Delta_Z^2}-\Omega \Big), \\
\text{Im}\Pi_{33}^{scsc} &= -\pi A \frac{2\Delta_Z^2(\Omega^2-\Delta_Z^2)}{\Delta_R^3\sqrt{\Delta_R^2+\Delta_Z^2-\Omega^2}}  \Theta(\Omega-\Delta_Z)\\
&\hspace{2.5cm} \times \Theta \Big( \sqrt{\Delta_R^2+\Delta_Z^2}-\Omega \Big), \\
\text{Im}[i\Pi_{12}^{scsc}] &= \pi A \frac{2\Omega \Delta_Z \sqrt{\Delta_R^2+\Delta_Z^2-\Omega^2}}{\Delta_R^3}  \Theta(\Omega-\Delta_Z)\\
&\hspace{2.5cm} \times \Theta \Big( \sqrt{\Delta_R^2+\Delta_Z^2}-\Omega \Big).
\end{split}
\eeq
The conditions imposed by $\Theta$-functions in the above Eq.~\er{ImPi_Z} clearly impliy the range of the particle-hole continuum to be $\Delta_Z<\Omega<\sqrt{\Delta_R^2+\Delta_Z^2}$.

According to the easier way adopted for $\Delta_Z=0$ case (more specifically Eq.~\er{trans}) in the previous section, the location of the edges of the particle-holes continuum at finite $\Delta_Z$ can be obtained from the single-particle dispersion:
\beq
\label{trans_Z}
\hbar \Omega = \ve_+ - \ve_- = \sqrt{\Delta_R^2 \sin^2\theta + \Delta_Z^2}.
\eeq
Since $\sin^2\theta$ is bounded between 0 and 1, the minimum and maximum values of $\Omega$ are $\Delta_Z$ and $\sqrt{\Delta_R^2 + \Delta_Z^2}$, respectively. These are identified as lower and upper edges of the continuum, respectively, which matches with the one obtained from an explicit calculation in Eq.~\er{ImPi_Z}.

The main difference between $\Delta_Z=0$ and finite $\Delta_Z$ cases is that in the latter case the lower edge of the particle-hole continuum is gapped. This indicates that in addition to collective modes above the continuum, the system may also support collective modes below the continuum. Indeed, the divergence in the real part of $\Pi_{ij}^{scsc}$ at both $\Omega=\Delta_Z$ (lower edge) and $\Omega=\sqrt{\Delta_R^2+\Delta_Z^2}$ (upper edge) ensures poles in the full phonon propagator both below and above the particle-hole continuum.

Let us now comment on the role of the field direction.
The opening of the low-energy gap in the continuum at low fields is unique to magnetic field oriented along the polar axis.
If compared with the situation when $\textbf{B} \perp \bP_0$, we find that the gap below the continuum does not open at weak fields. More specifically, when $\Delta_Z<\Delta_R$, the continuum is gapless. The gap opens up only when $\Delta_Z>\Delta_R$ with the lower edge of the continuum located at $(\Delta_Z-\Delta_R)$. To see this, let us assume $\textbf{B} \parallel \hat{x}$ and the local polar order still aligned with the $z$-axis. The $\textbf{B} \perp \bP_0$ analogue of Eq.~\er{trans_Z} can be calculated as
\beq
\label{perp}
\hbar \Omega = \ve_+ - \ve_- = \sqrt{\Delta_R^2 \sin^2\theta + \Delta_Z^2 - 2\Delta_R \Delta_Z \sin\theta\sin\phi}.
\eeq
For, $\Delta_Z<\Delta_R$, the minimum of Eq.~\er{perp} is at $\Omega=0$ which exists at $\phi=\pi/2$ and $\theta=\sin^{-1}(\Delta_Z/\Delta_R)$. The continuum, therefore, is gapless. For $\Delta_Z>\Delta_R$, however, the minimum shifts at $\Omega=\Delta_Z-\Delta_R$ which exists at $(\theta, \phi) = (\pi/2, \pi/2)$ resulting in the gapped continuum. Finally, the upper edge of the continuum lies at $\Omega=\Delta_Z+\Delta_R$. So, the uniqueness of $\textbf{B} \parallel \bP_0$ scenario is the continuum being gapped at low energies already at weak magnetic fields to support collective modes in it.

\subsection{Coupled lattice and electronic collective modes}
\label{CM}
The next step is to calculate collective modes of the system in the spin-sector. The determinant in Eq.~\er{det} suggests decoupling of the in-plane and out-of-plane sectors of the renormalized phonon propagator $\hat{\mathcal{D}}$. The resulting equations giving rise to collective modes in these sectors are, therefore, decoupled. We will discuss collective modes in these sectors first at $\Delta_Z=0$ in Sec.~\ref{sec:zero_field} followed by at finite $\Delta_Z$ in Sec.~\ref{finite_field}.

\subsubsection{Collective modes at $\Delta_Z=0$}
\label{sec:zero_field}
In the non-polar phase \cite{kumar_polar}, the phonon self-energy vanishes at $\Delta_Z=0$ and the collective modes are described by the hybridization of the electronic plasmon and phonons \cite{mahan:book,Ruhman:2016}. In the polar phase, however, the phonon self-energy is finite already at $\Delta_Z=0$ owing to the appearance of the Rashba spin-splitting. Therefore, here we focus on the high electronic density regime, where hybridization with plasmons (or, alternatively, LO-TO phonon splitting) can be neglected, to focus on the new effects of the coupling \er{coupling}.

We will now study the renormalized phonon propagator. $\Pi_{12}^{scsc}$ and $\Pi_{33}^{scsc}$ vanish identically at $\Delta_Z=0$ as discussed in Sec.~\ref{scsc_Z=0}.  
The only surviving component (as per all the other symmetry arguments applicable to our model as discussed in Sec.~\ref{RPA}), therefore, is $\Pi_{11}^{scsc}$, and from the rotational symmetry $\Pi_{22}^{scsc} = \Pi_{11}^{scsc}$. Equation~\er{det} then reduces to
\beq
\label{det1}
\begin{split}
\frac{\Omega_n^2 + \omega_\parallel^2}{\Omega_0^2} \Bigg[ \frac{\Omega_n^2 + \omega_\perp^2}{\Omega_0^2} - A \bigg( 2 -& \frac{\Omega_n^2}{\Delta_R \sqrt{\Omega_n^2 + \Delta_R^2}} L_1(\Omega_n) \bigg) \Bigg]^2 \\
&= 0,
\end{split}
\eeq
where $L_1(\Omega_n) = L(\Omega_n, \Delta_Z=0)$ \er{L}. 
\begin{figure*}
\centering
\includegraphics[scale=0.55]{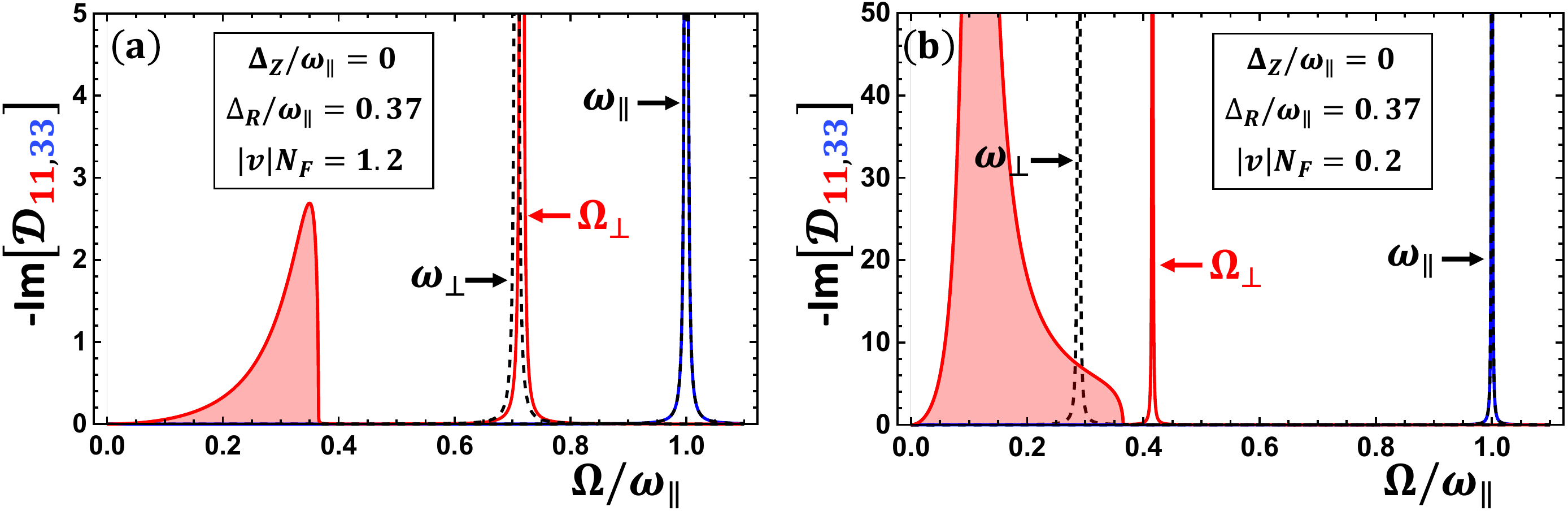}
\caption{\label{zero_field} Imaginary part of the renormalized phonon propagator at $\Delta_Z=0$ for cases when $\omega_\perp\gtrsim\Delta_R$ (a) and $\omega_\perp\lesssim \Delta_R$ (b). The model parameters are chosen as $\Omega_0=1.2\omega_\parallel$, $gN_F=2.4$ and $A=0.02$. The color scheme is the same as in Fig. \ref{fig:pi_scsc}, while dashed black lines correspond to the bare phonon in the absence of interactions. The shaded region in red represents the particle-hole continuum which ranges between $0 < \Omega < \Delta_R$. Sharp peaks in blue and red are modes polarized along the direction of the polar order $z$ (non-degenerate) and in the plane perpendicular to it (double-degenerate), respectively. 
In (a), two peaks above the continuum correspond to phonon-like modes, where only $\omega_\perp$ is renormalized due to interactions. In (b), the bare phonon frequency is within the continuum, apparently leading to its Fano shape at low energies. At the same time, a peak emerges above $\Delta_R$ that can be attributed to an electronic collective mode driven by divergence of the electronic response at $\Delta_R$ (see Fig.~\ref{fig:pi_scsc}(a)).}
\end{figure*}
After analytical continuation, Eq.~\er{det1} yields two positive definite roots, which are shown in Figs.~\ref{zero_field}(a) and \ref{zero_field}(b) as sharp red and blue peaks above the shaded region. The shaded region is nothing but the spin-flip particle-hole continuum as discussed in Sec.~\ref{scsc_Z=0}. One can verify that the continuum response is present in the region $0<\Omega<\Delta_R$ as consistent from our discussion in Sec.~\ref{scsc_Z=0}. The black dashed peaks in Figs.~\ref{zero_field}(a) and \ref{zero_field}(b) are bare phonons ($\Omega=\omega_\perp$ and $\Omega=\omega_\parallel$) that one expects in the absence of interaction. Moreover, Eq.~\er{det1} suggests that the solution $\Omega=\omega_\perp$ (in the absence of interaction) is double-degenerate. Physically, this is due to the in-plane (x-y) rotational symmetry of the system.

As one can see in Fig.~\ref{zero_field}, and also in Eq.~\er{det1}, that in the presence of interaction one of the modes still remains bare: the blue peak above the continuum at $\Omega=\omega_\parallel$, corresponding to the mode polarized along the direction of the polarization vector, coincides with the bare dashed-black peak. This is due to the fact that all the \{z\}-sector components of the self-energy, $\Pi_{3i}$, $\forall$ $i \in 1..3$, vanish at $\Delta_Z=0$, as discussed in Sec.~\ref{scsc_Z=0}. Therefore, the mode in the \{z\}-sector remains protected from renormalization due to interaction.

The double-degenerate mode, on the other hand, is polarized in the plane (x-y) perpendicular to the polarization vector and renormalizes due to interactions, as shown by red peaks above the continuum in Figs.~\ref{zero_field}(a) and \ref{zero_field}(b).

Before getting into the details of collective modes, we first discuss qualitative features of the single-particle continuum which evolves due to its interaction with phonons. This is evident from the difference in the shape of the continuum in Figs.~\ref{zero_field}(a) and \ref{zero_field}(b) as the anisotropy parameter in the Lagrangian \er{fe}, ``$v$", changes.

Degrees of freedom corresponding to electrons and phonons are decoupled in the absence of interactions: while the imaginary part of $\mathcal{D}_{ii}^0$ exhibits sharp phonon peaks, those of $\Pi_{11}^{scsc}$ exhibits broad continuum ranging from $0<\Omega<\Delta_R$ with a square-root divergence at $\Omega = \Delta_R$ \er{im11}. In the presence of interaction two things happen: first, the divergence in Im$\Pi_{11}^{scsc}$ at $\Omega=\Delta_R$ turns into zero of the full phonon propagator according to Eq.~\er{full_ph} and, second, the continuum evolves into a Fano-shaped resonance, as presented in Figs.~\ref{zero_field}(a) and \ref{zero_field}(b). To demonstrate this, we start with a situation when phonons are far above the continuum, or $\Delta_R \ll \omega_\perp \lesssim \omega_\parallel$, as shown in Fig.~\ref{zero_field}(a). At the level of the choice of parameters, this can be achieved by choosing a larger value of $|v|$. 
As the value of $v$ is decreased such that $\omega_\perp$ lies inside the continuum, i.e. $\omega_\perp < \Delta_R$, the spectral weight of the continuum increases demonstrating a Fano-like resonance feature; see Fig.~\ref{zero_field}(b). This suggests that interaction of the phonon with the continuum turn the typical Lorentzian shape of the phonon peak into a Fano-like one. However, there still remains a well-pronounced Lorentzian peak above the continuum edge.

We now discuss collective modes, the localized resonances above the continuum edge. As mentioned earlier in this section, mode polarized along the direction of the polar order ($\omega_\parallel$) remains unaffected from electron-phonon interaction \er{coupling}. The reason behind this is the conservation of $(\bk \times \hat{\bs})_z$ as explained in Sec.~\ref{scsc_Z=0}. On the other hand, the double-degenerate mode at $\omega_\perp$ renormalizes due to interaction, as represented by a shift in the red peak of Fig.~\ref{zero_field} from its non-interacting counterpart (dashed-black), and exists above the single-particle continuum. To highlight the splitting of modes from the particle-hole continuum, we chose an very small value of the damping parameter, $\gamma=10^{-4}\omega_\parallel$, in both the panels of Fig.~\ref{zero_field}. The same value of the damping parameter will be considered for later figures as well, i.e., while discussing collective modes at finite $\Delta_Z$, Fig.~\ref{fig:Z}, and electronic spin response effects at $\Delta_Z=0$ and $\Delta_Z\neq0$, Fig.~\ref{fig:chi}.

To  gain analytic insight, we search for the solution of Eq.~\er{det1} (second term) close to the upper edge of the continuum, i.e., $\Omega \approx \Delta_R + E_B$, with $0 < E_B \ll \Delta_R$. Here $\Omega$ is the retarded frequency, $E_B$ is the binding energy, and $\Delta_R$ \er{rash} is the Rashba energy. Upon using Eq.~\er{L} to write $L_1(\Omega_n)$, we get the required equation (second term of Eq.~\er{det1}) to be solved:
\beq
\label{det2}
\begin{split}
\frac{\omega_\perp^2 - \Delta_R^2}{\Omega_0^2} - A \Bigg( 2 + \frac{\Delta_R}{\sqrt{-2E_B\Delta_R}} &\text{log} \bigg[ \frac{\sqrt{-2E_B\Delta_R} + \Delta_R}{\sqrt{-2E_B\Delta_R} - \Delta_R} \bigg] \Bigg) \\
&= 0.
\end{split}
\eeq
Now expanding Eq.~\er{det2} for small $E_B$, and then upon solving for it, we get
\beq
\label{freq1}
\Omega_\perp \approx \Delta_R \bigg( 1 + \frac{A^2 \pi^2 \Omega_0^4}{2(\Delta_R^2 - \omega_\perp^2 + 4A\Omega_0^2)^2} \bigg).
\eeq
The binding energy is the second term of Eq.~\er{freq1}. 

It is essential to know the regime of validity of Eq.~\er{freq1}. Since this solution is valid when $E_B \ll \Delta_R$, let's consider only the second term of Eq.~\er{det2}, and expand it for small $E_B$:
\beq
\label{2nd}
2^\text{nd} \text{term} \approx -A \bigg( -\pi \sqrt{\frac{\Delta_R}{2 E_B}} + 4 + \mathcal{O}(E_B) \bigg).
\eeq
From the first two terms, it is easy to deduce that the above expansion is valid only when $E_B < (\pi^2/32) \Delta_R$. Moreover, the overall sign of \er{2nd} is positive, which indicates that Eq.~\er{det2} yields a real solution only when $\Delta_R > \omega_\perp$: while the second term of Eq.~\er{det2} is positive for $E_B < (\pi^2/32) \Delta_R$ (upon taking into account the overall minus sign in Eq.~\er{2nd}), the first term has to be negative for any real solution to exist. The regime of validity ($\Delta_R > \omega_\perp$) of the solution in Eq.~\er{freq1} suggests it be best represented by Fig.~\ref{zero_field}(b) when one of the phonons is inside the continuum, or $\Delta_R>\omega_\perp$.

Since the solution \er{freq1} obtained above is valid for $\Delta_R > \omega_\perp$, it is now imperative to know the fate of the collective mode in the opposite limit when $\Delta_R < \omega_\perp$. We look for solution near $\Omega \approx \omega_\perp$, which is expected to be valid when $\omega_\perp\gg\Delta_R$. For that, we consider Eq.~\er{det1} (its second term is relevant for the interaction-renormalized mode) and rewrite it after analytical continuation to real frequencies as
\beq
\label{36b}
\begin{split}
\frac{\omega_\perp^2 - \Omega^2}{\Omega_0^2} - A \bigg( 2 + \frac{\Omega^2}{\Delta_R \sqrt{\Delta_R^2 - \Omega^2}} \text{log} &\bigg[ \frac{\sqrt{\Delta_R^2 - \Omega^2} + \Delta_R}{\sqrt{\Delta_R^2 - \Omega^2} - \Delta_R} \bigg] \bigg) \\
&=0
\end{split}
\eeq
In the second term of Eq.~\er{36b}, we can replace $\Omega$ by $\omega_\perp$. This is reasonable because we are looking for a solution at $\Omega \approx \omega_\perp$. Now since we are in the $\Delta_R\ll\omega_\perp$ regime, one can expand the second term of Eq.~\er{36b} assuming large $\omega_\perp$. Upon doing all the above, we get
\beq
\label{freq2}
\Omega_\perp \approx \omega_\perp \bigg( 1 + \frac{2A \Delta_R^2 \Omega_0^2}{3 \omega_\perp^4} \bigg).
\eeq
In summary, the asymptotes of the renormalized mode is
\beq
\label{asymp1}
\Omega_\perp \approx 
\begin{cases}
	\Delta_R \bigg( 1 + A^2 \frac{\pi^2 \Omega_0^4}{2(\Delta_R^2 - \omega_\perp^2 + 4A\Omega_0^2)^2} \bigg), & \Delta_R \gtrsim \omega_\perp \\
	\omega_\perp \bigg( 1 + A\frac{2\Delta_R^2 \Omega_0^2}{3 \omega_\perp^4} \bigg).              & \Delta_R \ll \omega_\perp
\end{cases}
\eeq
One observes, that when the phonon energy is much greater than the continuum edge, the collective mode energy is very close to the phonon one, as shown in Fig.~\ref{zero_field}(a). However, when phonon energy is well within continuum, the collective mode closely follows the continuum edge, with a separation related to the electron-phonon coupling. This suggests that the origin of this mode is actually electronic, rather then lattice one, whereas the phonon contribution is observed in the Fano-shaped peak at low energies; see Fig.~\ref{zero_field}(b).

To summarize this section, at $\Delta_Z=0$, the mode polarized along the direction of the polar order (blue peak in Fig.~\ref{zero_field}) remains unaffected by electron-phonon interaction \er{coupling}. On the other hand, the double-degenerate mode, polarized in the plane perpendicular to the polar order (red peak in Fig.~\ref{zero_field}), is renormalized due to interaction. Moreover, a collective mode persists to occur even when the bare phonon energy is within the continuum as evidenced by the strong Fano peak. In this regime, its energy is very different from that of the phonon and can be interpreted as an electronic mode driven by electron-phonon interaction. Tuning the bare phonon frequency therefore interpolates between mostly electronic and mostly lattice origin of this mode.

\subsubsection{Collective modes at finite $\Delta_Z$}
\label{finite_field}
We will now consider finite Zeeman field case and discuss the fate of collective modes in polar metals due to it. Unlike at $\Delta_Z=0$ when the continuum is gapless, finite magnetic fields result in a gapped continuum ranging between $\Delta_Z<\Omega<\sqrt{\Delta_R^2 + \Delta_Z^2}$ (see Sec.~\ref{sec:se_Z}).

\begin{figure*}
\centering
\includegraphics[scale=0.54]{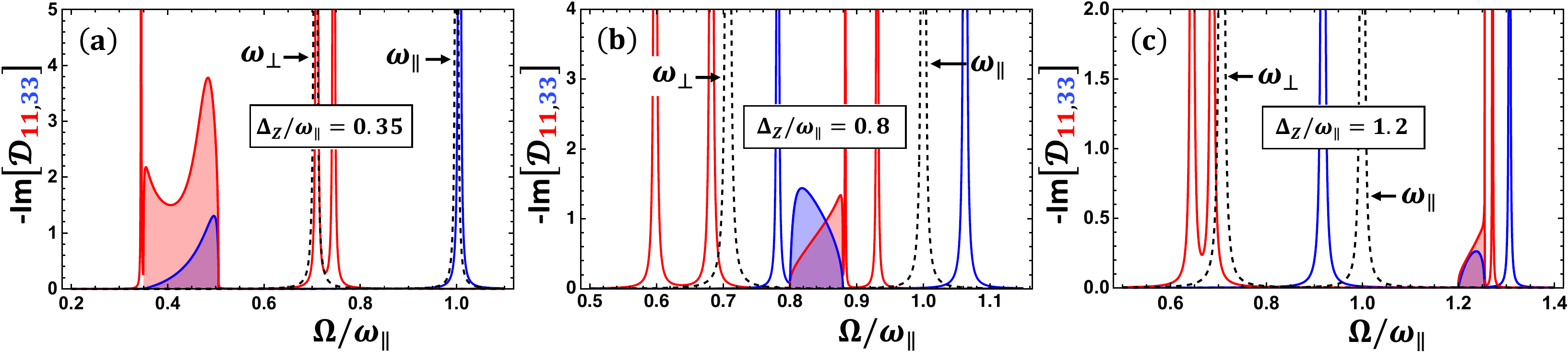}
\caption{\label{fig:Z} 
Imaginary part of the renormalized phonon propagator at different $\Delta_Z$ for fixed $\Delta_R=0.37\omega_\parallel$, $\Omega_0=1.2\omega_\parallel$, $gN_F=2.4$, $|v|N_F=1.2$ and $A=0.02$. The shaded regions in red and blue correspond to particle-hole spin-flip continuum, edges of which are located at $\Omega = \Delta_Z$ and $\Omega = \sqrt{\Delta_Z^2 + \Delta_R^2}$. Sharp peaks are resonances along the direction of polar order (blue) and in the plane perpendicular (red) to it. Dashed-black is the response of the bare phonons with $\omega_\perp < \omega_\parallel$.  
Increasing $\Delta_Z$ shifts the continuum energy such that (a) both phonons are above the continuum ($\sqrt{\Delta_R^2 + \Delta_Z^2} < \omega_\perp \lesssim \omega_\parallel$), (b) one above and one below the continuum ($\omega_\perp < \Delta_Z < \sqrt{\Delta_R^2 + \Delta_Z^2} < \omega_\parallel$), and (c) both below the continuum ($\omega_\perp \lesssim \omega_\parallel < \Delta_Z$).
In (a), the in-plane phonon-like peak around $\omega_\perp$ is split into two circularly-polarized modes. Below the continuum, an additional peak emerges 
which exists already at weak coupling because of the divergence in electronic response function at the edge of the continuum, see Fig. \ref{fig:pi_scsc}(b). In (c), the renormalized phonon response is below the continuum, while three electronic collective modes are seen to emerge above it. Importantly, a mode emerges also for the $zz$ polarization, as is expected from the divergence in the corresponding electronic response function, Fig. \ref{fig:pi_scsc}(b). (b) represents the case at intermediate fields: for the in-plane polarization the modes are similar to those in panel (c), whereas for the parallel polarization (along the direction of polar order which is $z$) an electronic mode appears below the continuum due to the continuum's sufficient proximity to $\omega_\parallel$ (see Eq.~\er{cond1} which is a condition for the appearance of this mode). On increasing the field (panel (c)), the character of the two modes in $zz$ reverses: electronic mode is above the continuum, while phonon-like one is below it.}
\end{figure*}

In Fig.~\ref{fig:Z} we show the imaginary part of the renormalized phonon propagator after continuation to real frequencies. The presence of the continuum can be noted from the red and blue shaded regions in Fig.~\ref{fig:Z}.
We note that unlike for $\Delta_Z=0$, the blue shaded region in Fig.~\ref{fig:Z} appears at finite $\Delta_Z$. This is because $\Pi_{33}^{scsc} \neq 0$ at finite fields which is not the case for $\Delta_Z=0$ where it is zero. More specifically, it's the Im$\Pi_{33}^{scsc} \neq 0$ which leads to the blue shaded continuum in the \{z\}-sector at finite $\Delta_Z$.
The divergence in Re$\Pi_{ij}^{scsc}$ at both the edges of the continuum (distributed between different components of the self-energy), as discussed in Sec.~\ref{sec:se_Z}, allows for the existence of collective modes both above and below the continuum (sharp red and blue peaks both below and above the continuum in Fig.~\ref{fig:Z}).

To calculate collective modes, we start with the main equation $\text{Det}[\mathcal{D}^{-1}] = 0$ \er{det}.
Using Eqs.~\er{phgr1} and \er{Pi_scsc}, Eq.~\er{det} can be written explicitly for in-plane and out-of-plane sectors as
\bwt
\bse
\beq
\label{eq1}
\text{out-of-plane sector}: -\frac{\Omega_n^2 + \omega_\parallel^2}{\Omega_0^2} + A \frac{\Delta_Z^2}{\Delta_R^2} \Bigg[ 4 - \frac{2 (\Omega_n^2 + \Delta_Z^2)}{\Delta_R \sqrt{\Omega_n^2 + \Delta_R^2 + \Delta_Z^2}} \text{log} \Bigg[ \frac{\sqrt{\Omega_n^2 + \Delta_R^2 + \Delta_Z^2} + \Delta_R}{\sqrt{\Omega_n^2 + \Delta_R^2 + \Delta_Z^2} - \Delta_R} \Bigg] \Bigg] = 0,
\eeq
\beq
\label{eq2}
\begin{split}
\text{in-plane sector}: \frac{\Omega_n^2 + \omega_\perp^2}{\Omega_0^2} - A \Bigg[ 2 &- \frac{4\Delta_Z^2}{\Delta_R^2} + \frac{\Delta_R^2 (-\Omega_n^2 + \Delta_Z^2) + 2\Delta_Z^2 (\Omega_n^2 + \Delta_Z^2)}{\Delta_R^3 \sqrt{\Omega_n^2 + \Delta_R^2 + \Delta_Z^2}} \text{log} \Bigg[ \frac{\sqrt{\Omega_n^2 + \Delta_R^2 + \Delta_Z^2} + \Delta_R}{\sqrt{\Omega_n^2 + \Delta_R^2 + \Delta_Z^2} - \Delta_R} \Bigg] \Bigg] \\
&= \mp A \frac{4i\Omega_n \Delta_Z}{\Delta_R^2} \Bigg[ 1 - \frac{\sqrt{\Omega_n^2 + \Delta_R^2 + \Delta_Z^2}}{2\Delta_R} \text{log} \Bigg[ \frac{\sqrt{\Omega_n^2 + \Delta_R^2 + \Delta_Z^2} + \Delta_R}{\sqrt{\Omega_n^2 + \Delta_R^2 + \Delta_Z^2} - \Delta_R} \Bigg] \Bigg],
\end{split}
\eeq
\ese
\ewt
where $A$ is the effective dimensionless coupling constant given in Eq.~\er{L}. Equations~\er{eq1} and \er{eq2} are transcendental in nature, so the exact analytical form of the solution cannot be obtained. Therefore, like in the previous section \ref{sec:zero_field}, we will provide analytical form of the solution only in asymptotic limits.

Although Eqs.~\er{eq1} and \er{eq2} are complicated, some crucial differences compared to those at $\Delta_Z=0$ \er{det1} can be recognized: (i) collective mode along the direction of polar order also renormalizes due to interaction as evident from Eq.~\er{eq1}, and (ii) the degeneracy of the mode polarized in the plane perpendicular to the polar order lifts due to time-reversal symmetry breaking as ensured by the $\pm$ sign (degeneracy lifting) which comes along with an overall factor of ``$i\Omega_n\Delta_Z$" (time-reversal symmetry breaking due to Zeeman fields) on the right hand side of Eq.~\er{eq2}. 

The poles of the phonon propagator correspond to solutions of these two equations and reveal themselves as Lorentzian peaks away from the continuum in Fig.~\ref{fig:Z}. 
In panels (a), (b) and (c) of Fig.~\ref{fig:Z} we present the imaginary part of the phonon propagator fro different values of $\Delta_Z$. Changing $\Delta_Z$ moves the boundaries of the continuum such that the collective modes change their positions with respect to the continuum, and new ones may appear. We note that it is different from the $\Delta_Z=0$ case, where the energy range of the continuum was fixed. There, $\omega_\perp$ (through $|v|$, as shown in Fig.~\ref{zero_field}) was used as a changing parameter, as discussed in Sec.~\ref{sec:zero_field}. At finite $\Delta_Z$,
for a given choice of parameters, the system hosts up to six collective modes (red and blue peaks in Figs.~\ref{fig:Z}(b) and (c)) which lie above and below the continuum. Equation~\er{eq1} gives up to two modes: one above the continuum and depending upon the value of $\Delta_Z$ one below the continuum, as presented by blue peaks in Figs.~\ref{fig:Z}(b) and (c). On the other hand, Eq.~\er{eq2} gives up to four modes: two modes above and up to two modes below the continuum, as presented by red peaks in Figs.~\ref{fig:Z}(b) and (c). 

Qualitatively, one may expect up to six different modes to exist. There are three phonons, polarized along three cartesian directions. Three electronic modes can be expected to arise from collective oscillations of spin along three cartesian directions. This is similar to as predicted in 2D electron systems with Rashba SOC and in-plane magnetic fields, \cite{maiti2016, kumar2017}. 

Let us now analyze the number of modes and their energies analytically. We will first discuss asymptotic expressions for the energy of collective modes below the continuum, Sec.~\ref{sec:below}, followed by those above the continuum, Sec.~\ref{sec:above}.

\paragraph{\textbf{Collective modes below the continuum}}
\label{sec:below}
To demonstrate collective modes below the continuum, we  consider three different cases corresponding to different $\Delta_Z$: (i) both $\omega_\perp$ and $\omega_\parallel$ above the continuum ($\Delta_Z<\sqrt{\Delta_Z^2+\Delta_R^2}<\omega_\perp<\omega_\parallel$), (ii) $\omega_\perp$ below and $\omega_\parallel$ above the continuum ($\omega_\perp < \Delta_Z < \sqrt{\Delta_R^2 + \Delta_Z^2} < \omega_\parallel$), and (iii) both $\omega_\perp$ and $\omega_\parallel$ below the continuum ($\omega_\perp \lesssim \omega_\parallel < \Delta_Z$), as illustrated in panels (a), (b) and (c) of Fig.~\ref{fig:Z}, respectively. This is achieved by fixing $\omega_\perp$, $\omega_\parallel$ and $\Delta_R$ and then varying $\Delta_Z$: $\omega_\parallel$ is already considered a fixed parameter, so we rescale all our energies by this; by fixing $v$ and $g$, and knowing that $\omega_\parallel$ is already fixed, we fix $\omega_\perp$ \er{phgr1}; finally, using $v$, $g$ and $\omega_\parallel$ as fixed parameters, we additionally fix the effective coupling constant $A$ to fix $\Delta_R$ \er{rash} as well. It is reasonable to fix $A$ assuming weak coupling ($A\ll1$) at which our theory is valid.

Assuming $\Delta_R$ and both the phonons, $\omega_\perp$ and $\omega_\parallel$, as fixed parameters, we found that there exist thresholds for $\Delta_Z$ (the threshold values for in-plane and out-of-plane modes are different which we will discuss below in this section) which guide the number of modes below the continuum in the in-plane and the out-of-plane sectors. For e.g., Fig.~\ref{fig:Z}(a) shows only one mode (red peak) below the continuum for small $\Delta_Z$. On the other hand, Figs.~\ref{fig:Z}(b) and (c) show three modes (red and blue peaks combined) below the continuum as $\Delta_Z$ is increased above a certain value. 
This deserves explanation which we provide by splitting the discussion of collective modes in the out-of-plane sector and in the in-plane sector.

\subparagraph{Modes in the out-of-plane-sector}
As we can see, in Fig.~\ref{fig:Z}(a) the collective mode in the \{z\}-sector (blue peak) is not present at small $\Delta_Z$, while it is present in Figs.~\ref{fig:Z}(b) and (c) when $\Delta_Z$ is above a certain value. To obtain the condition under which this happens we start with Eq.~\er{eq1} and look for mode just below the continuum: $\Omega \approx \Delta_Z - E_B$, where $E_B \ll \Delta_Z$ as the binding energy.
The log-divergence that seems to appear at $\Omega = \Delta_Z$ in Eq.~\er{eq1} disappears because of the prefactor $(\Omega_n^2 + \Delta_Z^2)$. So, as a first approximation we ignore this term completely. The dynamics is then governed by the first term of Eq.~\er{eq1} only. After analytical continuation to real frequencies, we replace $\Omega^2$ by $(-2\Delta_Z E_B + \Delta_Z^2)$ and solve for $E_B$. The collective mode just below the continuum can be calculated: 
\beq
\label{outbelow1}
\Omega_\parallel \approx \Delta_Z \Bigg( 1 - \frac{\Delta_R^2(\Delta_Z^2 - \omega_\parallel^2) + 4A\Omega_0^2 \Delta_Z^2}{2\Delta_R^2 \Delta_Z^2} \Bigg),
\eeq
where the binding energy can be read off as
\beq
\label{bind}
E_B = \frac{\Delta_R^2(\Delta_Z^2 - \omega_\parallel^2) + 4A\Omega_0^2 \Delta_Z^2}{2\Delta_R^2 \Delta_Z}.
\eeq
If the system remains stable, then the only reason we miss the mode (blue peak) in the \{z\}-sector, as apparent in Fig.~\ref{fig:Z}(a), is when the mode is Landau damped, or it merges with the continuum. This is possible when $E_B<0$, which ensures no pole below the continuum in this sector. Upon applying this, we get the condition:
\beq
\label{cond1}
\begin{split}
\Delta_Z &< \Delta_Z^{1c}, \\
\text{where} \,\, \Delta_Z^{1c} &\equiv \frac{\omega_\parallel \Delta_R}{\sqrt{\Delta_R^2 + 4A\Omega_0^2}}.
\end{split}
\eeq
So, for given $\Delta_R$, $\omega_\parallel$ and $\Omega_0$ ($A$ is fixed already while fixing $\Delta_R$), the system does not support any collective mode below the continuum in the \{z\}-sector when $\Delta_Z$ is small. More specifically, if $\Delta_Z < \Delta_Z^{1c}$, the mode is Landau damped, as evident from Fig.~\ref{fig:Z}(a) where the blue peak below the continuum is missing. The tendency of the \{z\}-sector mode to merge with the continuum is exactly due of the fact that the divergence in the Re$\Pi_{33}^{scsc}$ vanishes at the lower edge of the continuum ($\Omega = \Delta_Z$), as discussed in Sec.~\ref{sec:se_Z}.

The mode peels off the continuum from below only when $\Delta_Z$ is large enough, or $\Delta_Z>\Delta_Z^{1c}$; see blue peaks below the continuum in Figs.~\ref{fig:Z}(b) and (c). 
It is clear from Eq.~\er{cond1} that $\omega_\parallel > \Delta_Z^{1c}$ for $A>0$ and $\Delta_R>0$. Assuming that the collective mode below the continuum now exist in the \{z\}-sector, we have two possibilities of accommodating $\Delta_Z$ in this regime: $\Delta_Z^{1c}<\Delta_Z\lesssim\omega_\parallel$ and $\Delta_Z^{1c}<\omega_\parallel \ll \Delta_Z$. In other words, these regimes correspond to cases when the phonon polarized along the polar order ($\omega_\parallel$) is just above the continuum (specifically this means $\sqrt{\Delta_Z^2+\Delta_R^2}<\omega_\parallel$, but for the purpose of our discussion $\Delta_Z<\omega_\parallel$ assumption is sufficient) and when it is well below the continuum, as illustrated in Figs.~\ref{fig:Z}(b) and (c), respectively.

The former regime ($\Delta_Z^{1c}<\Delta_Z\lesssim\omega_\parallel$) hosts the solution obtained in Eq.~\er{outbelow1}, shown by blue peak just below the continuum in Fig.~\ref{fig:Z}(b). This mode is primarily electronic in nature. The reason is that the continuum interacts strongly with $\omega_\parallel$ as it moves towards it due to the increase in $\Delta_Z$. Due to this interaction, the continuum changes shape and skews away from $\omega_\parallel$ (notice this change in the blue shaded regions of Fig.~\ref{fig:Z}(b) as compared to that of Fig.~\ref{fig:Z}(a)) as the distance between them decreases. At some point, it leaves behind a finite spectral weight that closely follows it, as shown by sharp blue peak below the continuum in Fig.~\ref{fig:Z}(a). Since, this mode is split off the electronic continuum, the nature of it is primarily electronic.

The regime of validity of solution \er{outbelow1} is a bit subtle. As mentioned in the beginning of this section is that Eq.~\er{outbelow1} is valid only when $E_B \ll \Delta_Z$. 
Imposing this constraint, we get a condition: $\Delta_Z^2(\Delta_R^2-4A\Omega_0^2) \gg -\omega_\parallel^2\Delta_R^2$. 
If $\Delta_R\gg2\sqrt{A}\Omega_0$, then $E_B$ is always less than $\Delta_Z$ which is most likely the scenario at weak coupling ($A\ll1$). 
However, for sufficiently large $\Omega_0$ (e.g. $\Omega_0 \approx 194.4$ meV for SrTiO$_3$  \cite{yamada}), the case $\Delta_R < 2\sqrt{A}\Omega_0$ can be considered within the same formalism. 
Assuming this, we have a condition: 
\beq
\label{cond1b}
\begin{split}
\Delta_Z &\ll \Delta_Z^{2c},  \\
\text{where} \,\, \Delta_Z^{2c} &= \frac{\omega_\parallel\Delta_R}{\sqrt{4A\Omega_0^2-\Delta_R^2}}, \,\, \text{for} \,\,\Delta_R \lesssim 2\sqrt{A}\Omega_0.
\end{split}
\eeq
We note that for $\Delta_R \lesssim 2\sqrt{A}\Omega_0$, the regime of validity of solution \er{outbelow1} is $\Delta_Z^{1c}<\Delta_Z\ll\Delta_Z^{2c}$. 
To make our point, from now onwards, we always assume weak coupling, and also small enough $\Omega_0$, such that $\sqrt{A}\Omega_0 \ll \Delta_R$.

We now turn to the latter regime when $\omega_\parallel$ is well below the lower edge of the continuum ($\Delta_Z^{1c}<\omega_\parallel \ll \Delta_Z$), as illustrated in Fig.~\ref{fig:Z}(c).
The collective mode in this case can be interpreted as the phonon at $\omega_\parallel$ shifted in energy by the interaction.
To find the analytic form of this mode in the region deep below the continuum, similar to how it is shown in Fig.~\ref{fig:Z}(c) (blue peak below the continuum), we expand Eq.~\er{eq1} for large $\Delta_Z$ at $\Omega \approx \omega_\parallel$. 
Solving the resulting equation for $\Omega$, we get
\beq
\label{outbelow2a}
\begin{split}
\Omega_\parallel \approx \omega_\parallel \bigg( 1 - \frac{4 A \Omega_0^2}{3 \omega_\parallel^2} + \frac{4 A \Omega_0^2 (4\Delta_R^2-5\omega_\parallel^2)}{15 \omega_\parallel^2 \Delta_Z^2} + ... \bigg).
\end{split}
\eeq
From the above equation, one can easily write the collective mode when $\omega_\parallel$ and $\Delta_R$ soften near phase transition.
So far no such assumption is made so it is legitimate to say that the above solution \er{outbelow2a} is valid in the regime $\Delta_Z^{1c}<\omega_\parallel \sim \Delta_R \ll \Delta_Z$.

So, to conclude this section, in the out-of-plane sector, the system supports a collective mode below the continuum only when $\Delta_Z^{1c}<\Delta_Z$ \er{cond1}. 
The mode is primarily electronic, and given by Eq.~\er{outbelow1}, as long as $\Delta_Z^{1c}<\Delta_Z \lesssim \omega_\parallel$ for $\Delta_R\gg2\sqrt{A}\Omega_0$, or $\Delta_Z^{1c}<\Delta_Z\ll\Delta_Z^{2c}$ for $\Delta_R\lesssim2\sqrt{A}\Omega_0$; see blue peak below the continuum in Fig.~\ref{fig:Z}(c). 
At large magnetic fields when $\Delta_Z^{1c}<\omega_\parallel \ll \Delta_Z$, the collective mode evolves into a renormalized phonon as given by Eq.~\er{outbelow2a}. 
This is shown by blue peak below the continuum in Fig.~\ref{fig:Z}(b). The result of this section can be summarized as
\bwt
\beq
\label{asymp2}
    \Omega_\parallel \approx 
\begin{cases}
    \text{mode is Landau damped}, & \Delta_Z<\Delta_Z^{1c}\ll\omega_\parallel \\
    \Delta_Z \bigg( 1 - \frac{\Delta_R^2(\Delta_Z^2 - \omega_\parallel^2) + 4A\Omega_0^2 \Delta_Z^2}{2\Delta_R^2 \Delta_Z^2} \bigg), & \Delta_Z^{1c}<\Delta_Z \lesssim \omega_\parallel \,\, (\text{for} \,\, \Delta_R\gg2\sqrt{A}\Omega_0) \\
    & \,\, \text{or} \,\, \Delta_Z^{1c}<\Delta_Z\ll\Delta_Z^{2c} \,\, (\text{for} \,\, \Delta_R\lesssim2\sqrt{A}\Omega_0) \\
    \omega_\parallel \bigg( 1 - \frac{4 A \Omega_0^2}{3 \omega_\parallel^2} + \frac{4 A \Omega_0^2 (4\Delta_R^2-5\omega_\parallel^2)}{15 \omega_\parallel^2 \Delta_Z^2} + ... \bigg),              & \Delta_Z^{1c}<\omega_\parallel \sim \Delta_R \ll \Delta_Z
\end{cases}
\eeq
\ewt
where $\Delta_Z^{1c}$ and $\Delta_Z^{2c}$ are given by Eqs.~\er{cond1} and \er{cond1b}, respectively. In the crossover regime, the analytical form of the solution is not possible and we resort to numerics for the full solution which is presented by blue peaks below the continuum in Figs.~\ref{fig:Z}(b) and (c).

\subparagraph{Modes in the in-plane-sector} We now discuss collective modes in the in-plane, or \{x-y\}-, sector. The full spectrum of collective modes in this sector is shown by red peaks below the continuum in Figs.~\ref{fig:Z}(a), (b) and (c).

As shown in Fig.~\ref{fig:Z}, unlike in the \{z\}-sector, the system supports at least one mode below the continuum in the \{x-y\}-sector when both the phonons $\omega_\perp$ and $\omega_\parallel$ are above the continuum; 
for the purpose of this section, only $\omega_\perp$ is relevant as this phonon is polarized in the (x-y)-plane. 
This mode is purely electronic in nature, since its energy is clearly distinct from one of the phonons.
The simple reason behind this is the presence of log-divergence in both $\Pi_{11}^{scsc}$ and $\Pi_{12}^{scsc}$ at the lower edge of the continuum ($\Omega = \Delta_Z$) as discussed in Sec.~\ref{sec:se_Z}, given by Eq.~\er{Pi_scsc}, and shown in Fig.~\ref{fig:pi_scsc}(b). 
This is in contrast to $\Pi_{33}^{scsc}$ where the divergence at $\Omega = \Delta_Z$ vanishes. This poses a constraint on the existence of collective mode. Indeed, the collective mode in the \{z\}-sector appears only when $\Delta_Z$ is above a certain critical value ($\Delta_Z > \Delta_Z^{1c}$ to be specific), as discussed in the previous section and summarized in Eq.~\er{asymp2}.

To calculate collective modes just below the continuum in the in-plane sector, we adopt the same approach as discussed in the context of the out-of-plane mode: $\Omega \approx \Delta_Z - E_B$, with $E_B \ll \Delta_Z$. We can write Eq.~\er{eq2} within this approximation and expand for small $E_B$:
\beq
\label{42b1}
\begin{split}
\frac{\omega_\perp^2 - \Delta_Z^2}{\Omega_0^2} + \frac{2\Delta_Z E_B}{\Omega_0^2} &- A \Bigg( 2 - \frac{4\Delta_Z^2}{\Delta_R^2} - \frac{2\Delta_Z^2}{\Delta_R^2} \text{log}\frac{\Delta_Z E_B}{2\Delta_R^2} \Bigg) \\
&\pm A \frac{4\Delta_Z^2}{\Delta_R^2} \Bigg( 1 + \frac{1}{2} \text{log}\frac{\Delta_Z E_B}{2\Delta_R^2} \Bigg) = 0
\end{split}
\eeq
As we can see, Eq.~\er{42b1} yields two solutions, each corresponding to equations with $+/-$ sign in the last term. This is a primary indication of up to two modes below the continuum in the in-plane sector as also observed in Figs.~\ref{fig:Z}(b) and (c). However, whether both these modes are long lived and well-split from the continuum is something that deserves explanation: while in Figs.~\ref{fig:Z}(b) and (c) indeed there are two modes below the continuum, in Fig.~\ref{fig:Z}(a) there is only one. We will prove below that one of these modes actually merges into the continuum at some critical value of the Zeeman field, or alternatively, any other quantity chosen for convenience. The other mode, however, is robust and exponentially bound to the continuum. It is the robust mode that is shown in Fig.~\ref{fig:Z}(a).

Let's start with Eq.~\er{42b1} with a ``+" sign in the last term. The second term of this equation (proportional to $E_B$) is small, so we ignore this. The resulting equation can be re-arranged and written as
\beq
\label{42b2}
\text{log} \frac{\Delta_Z E_B}{2\Delta_R^2} = - \bigg( 2 - \frac{\Delta_R^2}{2\Delta_Z^2} + \frac{\Delta_R^2(\omega_\perp^2 - \Delta_Z^2)}{4A \Omega_0^2 \Delta_Z^2} \bigg).
\eeq
The log on the LHS of \er{42b2} is negative because of small $E_B$; specifically, this happens for $E_B \ll 2\Delta_R^2/\Delta_Z$. For any real solution to exist, we must have RHS of \er{42b2} to be negative as well. Assuming this to be the case, we solve for $E_B$ and get a solution:
\beq
\label{exp}
\Omega_\perp^+ \approx \Delta_Z - \frac{2\Delta_R^2}{\Delta_Z} \text{Exp} \bigg[ - \bigg( 2 - \frac{\Delta_R^2}{2\Delta_Z^2} + \frac{\Delta_R^2(\omega_\perp^2 - \Delta_Z^2)}{4A \Omega_0^2 \Delta_Z^2} \bigg) \bigg].
\eeq
As we see, the binding energy ($E_B$) in Eq.~\er{exp} has an exponential factor which must be less than 1 for the solution to be valid. This is indeed the case when $A\ll1$, i.e. for weak coupling.
The solution \er{exp} is a robust electronic mode, as shown in Fig.~\ref{fig:Z}(a), which never merges into the continuum, unlike the one found in Eq.~\er{outbelow1} in the \{z\}-sector. It remains exponentially bound to the continuum from below. As discussed earlier in this section, and also in Sec.~\ref{sec:se_Z}, the main reason behind the existence of this robust mode is the log-divergence in $\Pi_{11}^{scsc}$ and $\Pi_{12}^{scsc}$ components of the phonon self-energy.

Now we discuss the regime of validity of solution \er{exp}. Clearly, the argument of the exponential has to be negative for solution \er{exp} to exist. This happens when
\beq
\label{cond_exp}
\begin{split}
\Delta_Z &< \Delta_Z^{3c}, \\
\text{where} \,\, 
\Delta_Z^{3c} &\equiv \Delta_R\sqrt{\frac{\omega_\perp^2-2A\Omega_0^2}{\Delta_R^2-8A\Omega_0^2}}, \\
\text{for all} \,\, & \Delta_R > 2\sqrt{2A}\Omega_0 \,\, \text{and} \,\, \omega_\perp>\sqrt{2A}\Omega_0.
\end{split}
\eeq
The conditions on $\Delta_R$ and $\omega_\perp$ are legitimate at weak coupling ($A\ll1$) which is what we assume in our model. A more careful analysis suggests that $\Delta_Z^{3c} \gtrless \omega_\perp$ for $\omega_\perp \gtrless \Delta_R/2$. So, as a conclusion, the systems supports a robust electronic mode just below the continuum in the in-plane sector (see red peak below the continuum in Fig.~\ref{fig:Z}(a)) for $\Delta_Z<\Delta_Z^{3c}<\omega_\perp<\Delta_R/2$, with $\sqrt{2A}\Omega_0<\omega_\perp<\Delta_R/2$, or for $\Delta_Z\sim\Delta_R/2<\omega_\perp<\Delta_Z^{3c}$, with $\sqrt{2A}\Omega_0<\Delta_R/2<\omega_\perp$. 
Note that at any rate the condition given in Eq.~\er{cond_exp} must be satisfied for the solution \er{exp} to exist.

As $\Delta_Z$ increases such that $\Delta_R/2<\omega_\perp \lesssim \Delta_Z<\Delta_Z^{3c}$, the mode evolves into the one with a mixed character of electron and phonon: as continuum moves closer to $\omega_\perp$ (this happens because $\Delta_Z$ increases), the interaction between them results in a finite spectral weight below the continuum of mixed character, as shown by the lower energy red peaks ($\Omega_\perp^+$) in Fig.~\ref{fig:Z}(b). The analytical expression of the collective mode in this regime cannot be obtained, so we leave it for the numerical solution as shown already in Fig.~\ref{fig:Z}(b). The other higher energy red peak is the second mode ($\Omega_\perp^-$) in this sector, which we will discuss below in this section.

When we increase $\Delta_Z$ further such that $\Delta_Z>\Delta_Z^{3c}$, the solution \er{exp} is no longer valid. In such regime, for analytical insight, we assume $\Delta_Z$ to be large enough such that $\omega_\perp$ is well below the continuum, or more specifically, $\Delta_Z^{3c} \gtrless \omega_\perp \ll \Delta_Z$. In this regime, the collective mode is simply interaction-renormalized phonon, as shown by the low energy red mode below the continuum in Fig.~\ref{fig:Z}(c).

To obtain the analytical form of the renormalized phonon, we search for solution in the limit $\Omega \approx \omega_\perp \ll \Delta_Z$. We consider Eq.~\er{eq2} (with ``$-$" sign on the RHS) and expand it for large $\Delta_Z$ assuming $\Omega \approx \omega_\perp$. We solve the resulting equation for $\Omega$ to get
\beq
\label{in3}
\Omega_\perp^+ \approx \omega_\perp \bigg( 1 - \frac{2A\Omega_0^2}{3\omega_\perp^2} - \frac{2A\Omega_0^2}{3\omega_\perp \Delta_Z} - \frac{2A\Omega_0^2 (3\Delta_R^2 + 5\omega_\perp^2)}{15\omega_\perp^2\Delta_Z^2} + ... \bigg).
\eeq
At this point, no relative scaling between $\omega_\perp$ and $\Delta_R$ is assumed, so our result is valid in the $\Omega \approx \omega_\perp \sim \Delta_R \lessgtr \Delta_Z^{3c} \ll \Delta_Z$ regime.

Finally, at weak coupling $A<<1$, the result for $\Omega_\perp^+$ can be summarized as
\bwt
\beq
\label{asymp3}
    \Omega_\perp^+ \approx 
\begin{cases}
    \Delta_Z - \frac{2\Delta_R^2}{\Delta_Z} \text{Exp} \bigg[ - \bigg( 2 - \frac{\Delta_R^2}{2\Delta_Z^2} + \frac{\Delta_R^2(\omega_\perp^2 - \Delta_Z^2)}{4A \Omega_0^2 \Delta_Z^2} \bigg) \bigg], & \Delta_Z<\Delta_Z^{3c}<\omega_\perp<\Delta_R/2, \,\, \text{with} \,\, \sqrt{2A}\Omega_0<\omega_\perp<\Delta_R/2 \\
    \text{Same as above},              & \Delta_Z\sim\Delta_R/2<\omega_\perp<\Delta_Z^{3c}, \,\, \text{with} \,\, \sqrt{2A}\Omega_0<\Delta_R/2<\omega_\perp \\
    \text{Mixed character but could be same as above},              & \Delta_R/2<\omega_\perp \lesssim \Delta_Z<\Delta_Z^{3c}, \,\, \text{with} \,\, \sqrt{2A}\Omega_0<\Delta_R/2<\omega_\perp \\
    \omega_\perp \bigg( 1 - \frac{2 A \Omega_0^2}{3 \omega_\perp^2} - \frac{2 A \Omega_0^2}{3 \omega_\perp \Delta_Z} - \frac{2A\Omega_0^2 (3\Delta_R^2 + 5\omega_\perp^2)}{15\omega_\perp^2\Delta_Z^2} + ... \bigg),              & \omega_\perp \sim \Delta_R \lessgtr \Delta_Z^{3c} \ll \Delta_Z \\
\end{cases}
\eeq
\ewt
where $\Delta_Z^{3c}$ is given by Eq.~\er{cond_exp}.

The other solution of Eq.~\er{42b1} (corresponding to the one with a ``$-$" sign in the last term) just below the continuum can be obtained in the same way. However, if we notice, the last term (with ``$-$" sign) of Eq.~\er{42b1} is cancelled exactly by the other two terms in the equation. Therefore, in order to calculate the binding energy ($E_B$), one cannot ignore the second term (the one proportional to $E_B$) of this equation, as opposed to what we did while obtaining \er{exp}. Taking all this into account, we get the expression of the second mode:
\beq
\label{in2}
\Omega_\perp^- \approx \Delta_Z - \bigg( \frac{\Delta_Z}{2} + \frac{A\Omega_0^2}{\Delta_Z} - \frac{\omega_\perp^2}{2\Delta_Z} \bigg).
\eeq
The condition for this mode to exist below the continuum is obtained by imposing $E_B > 0$, where $E_B$ is given by terms inside the parenthesis of Eq.~\er{in2}. The resulting condition is
\beq
\label{cond3}
\begin{split}
\Delta_Z &> \Delta_Z^{4c}, \\
\text{where} \,\, \Delta_Z^{4c} &\equiv \sqrt{\omega_\perp^2 - 2A\Omega_0^2}.
\end{split}
\eeq
We again assume weak coupling, and small enough $\Omega_0$, such that $\omega_\perp \gg \sqrt{2A} \Omega_0$, which means $\Delta_Z^{4c}$ \er{cond3} is real. The solution \er{in2} is valid as long as $E_B \ll \Delta_Z$ which is true for all $\Delta_Z$ given the condition $\omega_\perp \gg \sqrt{2A} \Omega_0$ is satisfied. If somehow $\Omega_0$ is large enough such that $\omega_\perp < \sqrt{2A} \Omega_0$, then $E_B$ is always positive. However, in this case, for the solution \er{in2} to be valid, i.e., $E_B \ll \Delta_Z$, we now must have the condition $\Delta_Z >\sqrt{2A\Omega_0^2 - \omega_\perp^2}$. In any case, it is legitimate to assume $\Omega_0$ to be small enough such that the condition \er{cond3} is always satisfied.

We note that, as contrary to the mode exponentially close to the continuum \er{exp}, the mode obtained in Eq.~\er{in2} is not robust and has a tendency to merge into the continuum at some point. The reason behind this is the exact cancellation of log-divergence in Eq.~\er{42b1} (the one with ``$-$'' sign) where the solution \er{in2} is followed from. Just to remind, Eq.~\er{42b1} is nothing but Eq.~\er{eq2} written at $\Omega \approx \Delta_Z$.

It is clear that $\omega_\perp>\Delta_Z^{4c}$. So, the mode is Landau damped when $\Delta_Z<\Delta_Z^{4c}<\omega_\perp$ and does not show up below the continuum. This can be seen as an absence of second red peak below the continuum in Fig.~\ref{fig:Z}(a). The red peak which is shown in Fig.~\ref{fig:Z}(a) is the one given by Eq.~\er{exp}. The second mode peels off the continuum from below when $\Delta_Z^{4c}<\Delta_Z<\omega_\perp$ and given by Eq.~\er{in2}. 

As $\Delta_Z$ increases such that the phonon $\omega_\perp$ is just below the continuum, or $\Delta_Z^{4c}<\omega_\perp\lesssim\Delta_Z$, we reproduce the scanario as demonstrated in Fig.~\ref{fig:Z}(b). The sharp red peak below the continuum at higher energy is the one which demonstrates the second mode ($\Omega_\perp^-$). The one at lower energy is the first mode ($\Omega_\perp^+$) obtained from Eq.~\er{42b1} with the ``+" sign.

Finally, at large $\Delta_Z$, when the phonon $\omega_\perp$ is well below the continuum, $\Delta_Z^{4c}<\omega_\perp\ll\Delta_Z$, the collective mode is mostly a renormalized phonon. The solution of Eq.~\er{42b1} (with ``$-$" sign) is obtained by expanding the same at large $\Delta_Z$ assuming $\Omega \approx \omega_\perp$. Additionally, we also assume $\omega_\perp \sim \Delta_R$. The solution obtained in this regime is
\beq
\label{in4}
\Omega_\perp^- \approx \omega_\perp \bigg( 1 - \frac{2A\Omega_0^2}{3\omega_\perp^2} + \frac{2A\Omega_0^2}{3\omega_\perp \Delta_Z} - \frac{2A\Omega_0^2 (3\Delta_R^2 + 5\omega_\perp^2)}{15\omega_\perp^2\Delta_Z^2} + ... \bigg),
\eeq
which is shown by the higher energy sharp red peak below the continuum in Fig.~\ref{fig:Z}(c).

The result for $\Omega_\perp^-$ can summarized for small enough $\sqrt{A}\Omega_0$ (more precisely for $\sqrt{2A}\Omega_0 \ll \omega_\perp$) as
\bwt
\beq
\label{asymp4}
    \Omega_\perp^- \approx 
\begin{cases}
    \text{Mode is Landau damped}, & \Delta_Z<\Delta_Z^{4c}\ll\omega_\perp \\
    \Delta_Z - \Big( \frac{\Delta_Z}{2} + \frac{A\Omega_0^2}{\Delta_Z} - \frac{\omega_\perp^2}{2\Delta_Z} \Big),              & \Delta_Z^{4c}<\Delta_Z\ll\omega_\perp \\
    \omega_\perp \bigg( 1 - \frac{2 A \Omega_0^2}{3 \omega_\perp^2} + \frac{2 A \Omega_0^2}{3 \omega_\perp \Delta_Z} - \frac{2A\Omega_0^2 (3\Delta_R^2 + 5\omega_\perp^2)}{15\omega_\perp^2\Delta_Z^2} + ... \bigg),              & \Delta_Z^{4c} \ll \omega_\perp \sim \Delta_R \ll \Delta_Z \\
\end{cases}
\eeq
\ewt
where $\Delta_Z^{4c}$ is given by Eq.~\er{cond3}. It can be checked from Eqs.~\er{asymp3} and \er{asymp4} that $\Omega_\perp^+<\Omega_\perp^-$. This, however, changes above the continuum and we will see in Sec.~\ref{sec:above} that above the continuum $\Omega_\perp^+>\Omega_\perp^-$. We remind the readers that the superscript $+/-$ in the definition of mode frequency mainly refers to the solution of Eq.~\er{eq2} with $-/+$ on the RHS.

A comparison between $\Delta_Z^{4c}$ \er{cond3} and $\Delta_Z^{3c}$ \er{cond_exp} suggests that $\Delta_Z^{4c}<\Delta_Z^{3c}\lessgtr \omega_\perp$, as long as $\sqrt{2A}\Omega_0<\omega_\perp\lessgtr\Delta_R/2$. According to Eqs.~\er{asymp3} and \er{asymp4}, we conclude that when $\Delta_Z<\Delta_Z^{4c}<\Delta_Z^{3c}\lessgtr \omega_\perp$, we expect to see only one collective electronic mode in the in-plane sector which is exponentially bound to the continuum from below as shown by a sharp red peak in Fig.~\ref{fig:Z}(a). For $\Delta_Z$ just above $\Delta_Z^{4c}$, another mode peels off below the continuum, and in this regime we have two electronic modes below the continuum in this sector. As $\Delta_Z$ increases further and lies in the regime $\Delta_Z^{4c}<\omega_\perp\lesssim\Delta_Z<\Delta_Z^{3c}$, with $\sqrt{2A}\Omega_0<\Delta_R/2<\omega_\perp$, both the modes possess mixed nature of electrons and phonons, as shown by red peaks below the continuum in Fig.~\ref{fig:Z}(b). Finally, for large $\Delta_Z$, or $\Delta_Z^{4c}<\Delta_Z^{3c} \lessgtr \omega_\perp\ll\Delta_Z$, modes are primarily phonon-like and shown by sharp red peaks below the continuum in Fig.~\ref{fig:Z}(c).

\paragraph{\textbf{Collective modes above the continuum}} 
\label{sec:above}
We now discuss collective modes above the continuum. At the upper edge of the continuum, $\Omega = \sqrt{\Delta_R^2 + \Delta_Z^2}$, the real part of both in-plane and out-of-plane components of the phonon self-energy, $\Pi_{11}^{scsc}$ and $\Pi_{33}^{scsc}$ \er{Pi_scsc}, have a square root divergence. This is an indication that the system may always support collective modes above the continuum at arbitrarily weak coupling in both \{x-y\}- and \{z\}-sectors. 
The existence of a mode above the continuum in the \{z\}-sector at arbitrarily weak coupling (or all $\Delta_Z$) is in stark contrast to that below the continuum: below the continuum, the mode exists only for $\Delta_Z>\Delta_Z^{1c}$ \er{cond1}, as discussed in Sec.~\ref{sec:below}.
Parenthetically, we note that the off-diagonal component of the self-energy, $\Pi_{12}^{scsc}$ \er{Pi_scsc}, does not have any divergence at the upper edge of the continuum. However, this does not pose any problem because since the in-plane components are coupled, the divergence in only the diagonal components of this sector, $\Pi_{ii}^{scsc}$, with $i = \{1,2\}$, already ensures collective modes above the continuum in the \{x-y\}-sector. 

Above the continuum, Eq.~\er{eq1} (equation for the \{z\}-sector mode) yields one mode, whereas Eq.~\er{eq2} (equation for the \{x-y\}-sector modes) yields two modes, as shown by one blue and two red peaks, respectively, in all the panels of Fig.~\ref{fig:Z}. We first discuss collective modes in the \{z\}-sector, followed by same in the \{x-y\}-sector.

\subparagraph{Modes in the out-of-plane sector} We start with Eq.~\er{eq1} and look for the solution just above the upper edge of the continuum: $\Omega \approx \sqrt{\Delta_R^2 + \Delta_Z^2} + E_B$, with $E_B \ll \sqrt{\Delta_R^2 + \Delta_Z^2}$. Using this assumption, Eq.~\er{eq1} can be written as
\beq
\label{det1_Z}
\begin{split}
\frac{\Delta_R^2 + \Delta_Z^2 - \omega_\parallel^2}{\Omega_0^2} &+ A \frac{\Delta_Z^2}{\Delta_R^2} \Bigg( 4 + \frac{2\Delta_R}{\sqrt{-2E_B} (\Delta_R^2 + \Delta_Z^2)^{1/4}} \times \\
\times \text{log} & \bigg[ \frac{\sqrt{-2E_B} (\Delta_R^2 + \Delta_Z^2)^{1/4} + \Delta_R}{\sqrt{-2E_B} (\Delta_R^2 + \Delta_Z^2)^{1/4} - \Delta_R} \bigg] \Bigg) = 0,
\end{split}
\eeq
which can be expanded further for small $E_B$:
\beq
\label{det1_ex}
\frac{\Delta_R^2 + \Delta_Z^2 - \omega_\parallel^2}{\Omega_0^2} \approx A \frac{\sqrt{2} \pi \Delta_Z^2}{\sqrt{E_B} \Delta_R (\Delta_R^2 + \Delta_Z^2)^{1/4}}.
\eeq
The RHS of Eq.~\er{det1_ex} is always positive for $A>0$ and $\Delta_R > 0$. Therefore, any real solution of Eq.~\er{det1_ex} requires $\sqrt{\Delta_R^2 + \Delta_Z^2} > \omega_\parallel$ in the LHS. Upon solving for $E_B$, the collective mode frequency above the continuum can be obtained:
\beq
\begin{split}
\Omega_\parallel \approx \sqrt{\Delta_R^2 + \Delta_Z^2} \Bigg[ 1 + \frac{2\pi^2 A^2 \Delta_Z^4 \Omega_0^4}{\Delta_R^2(\Delta_R^2 + \Delta_Z^2) (\Delta_R^2 + \Delta_Z^2 - \omega_\parallel^2)^2} \Bigg].
\end{split}
\eeq
This mode is primarily electronic and replicates the scenario as shown by a blue peak above the continuum in Fig.~\ref{fig:Z}(c). 

In the opposite limit $\sqrt{\Delta_R^2 + \Delta_Z^2} \ll \omega_\parallel$, the solution is primarily phonon-like with an interaction correction proportional to $A$. To obtain the analytical expression of the mode well above the continuum, we expand Eq.~\er{eq1} for large $\Omega \sim \omega_\parallel \gg \sqrt{\Delta_R^2+\Delta_Z^2}$. Upon solving the resulting equation for $\Omega$, we get
\beq
\Omega_\parallel \approx \omega_\parallel \bigg( 1 + \frac{4A \Delta_Z^2 \Omega_0^2}{3 \omega_\parallel^4} + ... \bigg), 
\eeq
which is illustrated well by a blue peak above the continuum in Figs.~\ref{fig:Z}(a) and (b) for different values of Zeeman field.

In summary, the asymptotes of the mode polarized along the direction of the polar order is
\bwt
\beq
\label{asymp5}
    \Omega_\parallel \approx 
\begin{cases}
    \sqrt{\Delta_R^2 + \Delta_Z^2} \Bigg[ 1 + \frac{2\pi^2 A^2 \Delta_Z^4 \Omega_0^4}{\Delta_R^2(\Delta_R^2 + \Delta_Z^2) (\Delta_R^2 + \Delta_Z^2 - \omega_\parallel^2)^2} \Bigg],     & \sqrt{\Delta_R^2 + \Delta_Z^2} \gtrsim \omega_\parallel \\
    \omega_\parallel \bigg( 1 + \frac{4A \Delta_Z^2 \Omega_0^2}{3 \omega_\parallel^4} + ... \bigg).     & \sqrt{\Delta_R^2 + \Delta_Z^2} \ll \omega_\parallel
\end{cases}
\eeq
\ewt

\subparagraph{Modes in the in-plane sector} To calculate collective modes in the in-plane sector we use the same procedure as applied for the out-of-plane sector mode: 
$\Omega \approx \sqrt{\Delta_R^2 + \Delta_Z^2} + E_B$, with $E_B \ll \sqrt{\Delta_R^2 + \Delta_Z^2}$, in Eq.~\er{eq2}. At small $E_B$, we expand Eq.~\er{eq2} and obtain the equation that is required to be solved:
\beq
\label{det2_ex}
\begin{split}
& -\frac{\Delta_R^2 + \Delta_Z^2 - \omega_\perp^2}{\Omega_0^2} + \frac{A \pi \Delta_R}{\sqrt{2E_B} (\Delta_R^2 + \Delta_Z^2)^{1/4}} \\
&\hspace{2cm} - \frac{2A(\Delta_R^2 - 2\Delta_Z^2 \mp 2\Delta_Z \sqrt{\Delta_R^2 + \Delta_Z^2})}{\Delta_R^2} = 0.
\end{split}
\eeq
Notice that there are two equations in Eq.~\er{det2_ex} which are distinguished by $\mp$ sign in the last term. The origin of the $\mp$ term is basically $\Pi_{12}^{scsc}$ which is non-zero because of broken time-reversal symmetry at finite $\Delta_Z$.

Let's now analyze Eq.~\er{det2_ex}. The one with a ``+" sign in the last term is overall positive (excluding the overall minus sign in front of it). Similarly, the first term of Eq.~\er{det2_ex} is also positive for $\sqrt{\Delta_R^2 + \Delta_Z^2} > \omega_\perp$ (excluding the overall minus sign). Given that the second term is positive (including the overall plus sign in front of it), we have a positive definite solution for $E_B$. On the other hand, the one with a ``$-$" sign in the last term of Eq.~\er{det2_ex} is not necessarily either positive or negative. However, it is proportional to the coupling constant $A$. So, given that the first term of Eq.~\er{det2_ex} is overall negative for $\sqrt{\Delta_R^2+\Delta_Z^2} > \omega_\perp$, we can say that at weak coupling ($A\ll1$), the combination of first and third terms of Eq.~\er{det2_ex} is overall negative. We then again have a positive definite solution for $E_B$.
Upon solving Eq.~\er{det2_ex} for $E_B$, the collective mode for $\sqrt{\Delta_R^2 + \Delta_Z^2} > \omega_\perp$ can be obtained:
\beq
\label{freq1_Z}
\begin{split}
\Omega_\perp^\pm \approx & \sqrt{\Delta_R^2 + \Delta_Z^2} \Bigg[ 1 + \frac{\pi^2 A^2 \Delta_R^2}{2(\Delta_R^2 + \Delta_Z^2)} \Bigg\{ \frac{\Delta_R^2 + \Delta_Z^2 - \omega_\perp^2}{\Omega_0^2} \\
&+ 2A \bigg( 1 - \frac{2\Delta_Z \big( \Delta_Z \pm \sqrt{\Delta_R^2 + \Delta_Z^2} \big)}{\Delta_R^2} \bigg) \Bigg\}^{-2} \Bigg].
\end{split}
\eeq
This solution is represented by two red peaks above the continuum in Figs.~\ref{fig:Z}(b) and (c).

To get the solution in the opposite regime, $\sqrt{\Delta_R^2 + \Delta_Z^2} \ll \omega_\perp$, we expand Eq.~\er{eq2} for large $\Omega \sim \omega_\perp \gg \sqrt{\Delta_R^2+\Delta_Z^2}$, and solve the resulting equation for $\Omega$. The collective modes read
\beq
\label{freq2_Z}
\Omega_\perp^\pm \approx \omega_\perp \bigg( 1 \pm \frac{2A\Omega_0^2 \Delta_Z}{3\omega_\perp^3} + \frac{2A\Omega_0^2(\Delta_R^2 + \Delta_Z^2)}{3\omega_\perp^4} \bigg),
\eeq
which are shown by red peaks above the continuum in Fig.~\ref{fig:Z}(a).

Finally, the asymptotes of the mode in the in-plane sector, or polarized in the plane perpendicular to the polar order, are summarized:
\bwt
\beq
\label{asymp6}
    \Omega_\perp^\pm \approx
\begin{cases}
    \sqrt{\Delta_R^2 + \Delta_Z^2} \Bigg[ 1 + \frac{\pi^2 A^2 \Delta_R^2}{2(\Delta_R^2 + \Delta_Z^2)} \Bigg\{ \frac{\Delta_R^2 + \Delta_Z^2 - \omega_\perp^2}{\Omega_0^2} + 2A \bigg( 1 - \frac{2\Delta_Z \big( \Delta_Z \pm \sqrt{\Delta_R^2 + \Delta_Z^2} \big)}{\Delta_R^2} \bigg) \Bigg\}^{-2} \Bigg],  & \sqrt{\Delta_R^2 + \Delta_Z^2} \gtrsim \omega_\perp \\
    \omega_\perp \bigg( 1 \pm \frac{2A\Omega_0^2 \Delta_Z}{3\omega_\perp^3} + \frac{2A\Omega_0^2(\Delta_R^2 + \Delta_Z^2)}{3\omega_\perp^4} \bigg).     & \sqrt{\Delta_R^2 + \Delta_Z^2} \ll \omega_\perp
\end{cases}
\eeq
\ewt
It can be verified that above the continuum $\Omega_\perp^+>\Omega_\perp^-$. This means that $\Omega_\perp^-$ is always closer to the continuum as compared to $\Omega_\perp^+$ both below and above the continuum.

Let us now summarize the entire Sec.~\ref{CM}. We summarized our results of collective modes at $\Delta_Z=0$ in Eq.~\er{asymp1}. At $\Delta_Z=0$, collective modes exist only above the particle-hole continuum. Moreover, only the mode polarized in the plane perpendicular to the polar order undergoes renormalization due to interaction. At finite $\Delta_Z$, however, the system supports collective modes both below and above the particle-hole continuum. Moreover, at finite $\Delta_Z$ collective modes polarized both along the polar order and perpendicular to it get renormalized due to interactions. Below the continuum, our results for $z$-direction polarized (along the polar order) collective mode are summarized in Eq.~\er{asymp2}, while that for the ones polarized in the xy-plane (perpendicular to the polar order) are summarized in Eqs.~\er{asymp3} and \er{asymp4}. Finally, above the continuum, the results for the $z$-direction polarized mode is summarized in Eq.~\er{asymp5}, while that for the xy-plane polarized mode is summarized in Eq.~\er{asymp6}.
we conclude that be it above or below the continuum, collective modes which are close to the continuum are primarily electronic, whereas the ones which are well above or well below the continuum are primarily phonon-like.

\section{Predictions for Electron/Electric-dipole Spin Resonance Measurements}
\label{ESR}

In this section we demonstrate that the collective modes predicted above will have signatures in the dynamical spin susceptibility, contributing to ESR and EDSR measurements (see Sec. \ref{RPA}): the EDSR response is related to the imaginary part of the spin-susceptibility (see Eq.~\er{rel}), so the resonance features in the latter show up at same frequencies as in the former.
The field-tuned collective modes identified in the renormalized phonon
responses will only be accessible in the dynamical spin susceptibility if there is coupling between the spin and the spin-current degrees
of freedom.  Here we calculate the spin susceptibility  
and check that it displays the same poles as those in the renormalized
phonon propagator. We show that these collective modes appear
as sharp resonances in the imaginary part of the spin susceptibility
both below and above the particle-hole continuum, and thus we provide
signatures for future spectroscopy experiments.

\begin{figure*}
\centering
\includegraphics[scale=0.54]{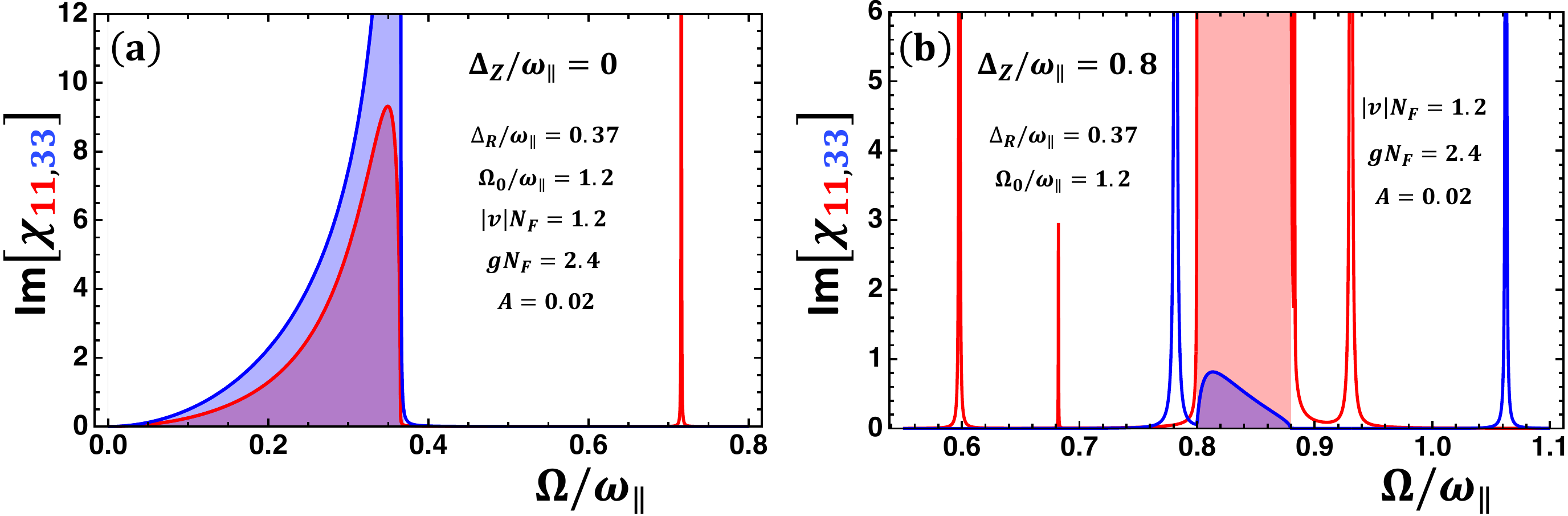}
\caption{\label{fig:chi} Imaginary part of the spin susceptibility, proportional to the ESR (and EDSR for the 22 component) intensity in polar metals at zero and finite magnetic fields for given parameter values. Shaded regions in blue and red
correspond to the spin-flip continuum, while sharp peaks outside the continuum are collective modes in the in-plane (red peaks) and out-of-plane (blue peaks) sectors. (a) Imaginary part of the spin-susceptibility (in units of $g^2\mu_B^2N_F/16$) at $\Delta_Z=0$. The peak above the continuum close to the bare phonon frequency indicates that the in-plane phonon becomes ESR active in the polar metal.
(b) Imaginary part of the spin-susceptibility (in units of $g^2\mu_B^2N_F/16$) at finite $\Delta_Z$. All the modes, in both in-plane (red peaks) and out-of-plane (blue peaks) sectors, that show up as poles in $\mathcal{D}_{ij}$ are ESR active. These modes show up as sharp resonances in Im$\chi_{ij}$ below and above the continuum.}
\end{figure*}

We calculate spin-susceptibility as defined in Sec.~\ref{RPA} and given by Eq.~\er{susc1}. For this, we require the form of spin-spin $(\Pi_{ij}^{ss})$ and spin - spin-current ($\Pi_{ij}^{ssc}$) correlation functions, which are defined by Eq.~\er{compact} in a compact form. The explicit form of coherence factors for these correlation functions, $f_{ij}^{r\bar{r}, ss}$ and $f_{ij}^{r\bar{r}, ssc}$, and the technical details of the calculation of bubble are delegated to Eqs.~\er{coh_ss} and \er{coh_ssc} of Appendix~\ref{appen:coh} and Appendix~\ref{appen:bubble}, respectively. Here we only provide final results. The explicit forms of non-zero $\Pi_{ij}^{ss}$ and $\Pi_{ij}^{ssc}$ can be calculated as
\beq
\label{Pi_ss}
\begin{split}
\Pi_{11}^{ss} (\Omega_n) &= - \frac{N_F}{4} \Bigg[ 2 + \frac{(-\Omega_n^2 + \Delta_Z^2)}{\Delta_R \sqrt{\Omega_n^2 + \Delta_R^2 + \Delta_Z^2}} L(\Omega_n) \Bigg], \\
\Pi_{22}^{ss} (\Omega_n) &= \Pi_{11}^{ss} (\Omega_n), \\
\Pi_{33}^{ss} (\Omega_n) &= - \frac{N_F}{4} \Bigg[ 4 - \frac{2(\Omega_n^2 + \Delta_Z^2)}{\Delta_R \sqrt{\Omega_n^2 + \Delta_R^2 + \Delta_Z^2}} L(\Omega_n) \Bigg], \\
\Pi_{12}^{ss} (\Omega_n) &= \frac{N_F}{4} \frac{2 \Omega_n \Delta_Z}{\Delta_R \sqrt{\Omega_n^2 + \Delta_R^2 + \Delta_Z^2}} L(\Omega_n), \\
\Pi_{21}^{ss} (\Omega_n) &= -\Pi_{12}^{ss} (\Omega_n),
\end{split}
\eeq
and
\beq
\label{Pi_ssc}
\begin{split}
\Pi_{33}^{ssc} (\Omega_n) &= \sqrt{A N_F} \frac{2\Delta_Z}{\Delta_R} \Bigg[ 1 - \frac{(\Omega_n^2 + 2\Delta_R^2 + \Delta_Z^2) L(\Omega_n)}{2\Delta_R \sqrt{\Omega_n^2 + \Delta_R^2 + \Delta_Z^2}} \Bigg], \\
\Pi_{11}^{ssc} (\Omega_n) &= \Pi_{22}^{ssc} (\Omega_n) = - \frac{1}{2} \Pi_{33}^{ssc} (\Omega_n), \\
\Pi_{12}^{ssc} (\Omega_n) &= - \frac{\Omega_n}{2\Delta_Z} \Pi_{33}^{ssc} (\Omega_n), \\
\Pi_{21}^{ssc} (\Omega_n) &= - \Pi_{12}^{ssc} (\Omega_n),
\end{split}
\eeq
respectively, where $N_F$ is the total density of states in 3D and, $A$ and $L(\Omega_n)$ are given by Eq.~\er{L}. As discussed earlier in Sec.~\ref{RPA}, the applicability of ESR/EDSR in the ordered phase depends on whether $\Pi_{ij}^{ssc}$ is finite or not. We emphasize that $\Pi_{ij}^{ssc}$ determines the spectral weight of the collective mode and is finite only in the ordered phase. To see this we expand $\Pi_{33}^{ssc}$ in Eq.~\er{Pi_ssc} for small $\Delta_R$:
\beq
\label{prop}
\Pi_{33}^{ssc} (\Omega_n) \approx \sqrt{A N_F} \bigg( - \frac{8 \Delta_Z \Delta_R}{3(\Omega_n^2 + \Delta_Z^2)} + \frac{8 \Delta_Z \Delta_R^3}{5(\Omega_n^2 + \Delta_Z^2)^2} + .... \bigg).
\eeq
Since $\Pi_{33}^{ssc} \propto \Delta_R$, it is finite only in the ordered phase according to Eqs.~\er{alpha} and \er{rash}. This proportionality \er{prop} holds true for other components of $\hat{\Pi}^{ssc}$ as well according to the relations mentioned in Eq.~\er{Pi_ssc}. If $\Pi_{ij}^{ssc}$ vanishes, which would be the case in the non-polar (paraelectric) phase where the static Rashba term \er{stat_rash} is zero, then the Im$\chi$ will have signal corresponding to only single-particle effects coming solely from the non-interacting spin polarization bubble, $\Pi_{ij}^{ss}$. There is no interaction effect that spin response would be able to capture in the non-polar phase.

In the next two sections, we will discuss spin response at zero \ref{ESR_zero} and finite \ref{ESR_finite} magnetic fields. Our aim is to study whether signatures of the collective modes discussed in Secs.~\ref{sec:zero_field} and \ref{finite_field} can appear in the spin-susceptibility. If so, then they can be resonantly excited in ESR by applying an ac magnetic field: in order to excite in-plane modes (either $x$ or $y$), one needs to apply an ac magnetic field along the $x$- or $y$-direction, the out-of-plane modes can be excited by applying an ac magnetic field along $z$-direction.

\subsection{Spin response without magnetic field}
\label{ESR_zero}
As discussed in Sec.~\ref{scsc_Z=0}, at zero magnetic field the particle-hole continuum is gapless and the collective modes exist only above the continuum, see Sec.~\ref{sec:zero_field} for details on collective modes at $\Delta_Z=0$. The mode in the \{z\}-sector does not renormalize due to interaction because the corresponding component of the self-energy \er{Pi_scsc} vanishes in this limit. Hence, we get bare phonons ($\omega_\parallel$) polarized along the $z$-direction. The mode in the \{x-y\}-sector, however, renormalizes due to interaction, the form of which is given by Eq.~\er{asymp1} in different regimes.

To know whether these modes are excited in ESR/EDSR, we calculate the corresponding observable, the spin-susceptibility, and see if same poles as in $\mathcal{D}_{ij}$ show up in the spin-susceptibility (see Eq.~\er{susc1} for the definition) as well or not. The result is shown in Fig.~\ref{fig:chi}(a) (for same parameter values as chosen in Fig.~\ref{zero_field}(a), including the damping parameter which is chosen to be $\gamma=10^{-4}\omega_\parallel$), where only the red peak above the continuum, corresponding to the mode in the \{x-y\}-sector, that appears in the imaginary part of the spin-susceptibility.
The location of the peak coincides with the one in Fig.~\ref{zero_field}(a). This is also clear from the definition of the spin-susceptibility in Eq.~\er{susc1}, where the in-plane components of renormalized phonon propagator appear in the in-plane components of the spin-susceptibility, while the out-of-plane component of the phonon propagator shows up in the out-of-plane component of the spin-susceptibility.

The spectral weight of the mode in the out-of-plane sector (blue peak) is zero and, therefore, is not excited in ESR/EDSR. This can be understood by looking at the explicit form of $\chi_{33} \er{susc1}$ which depends on $\Pi_{33}^{ssc}$ \er{Pi_ssc}. At $\Delta_Z=0$, $\Pi_{33}^{ssc}$ vanishes \er{Pi_ssc}, which in turn gives $\chi_{33} = \Pi_{33}^{ss}$. As we see, the out-of-plane susceptibility does not probe the pole in the $\mathcal{D}_{33}$ component; it only probes the single-particle continuum coming from Im$\Pi_{33}^{ss} \neq 0$, which is shown by blue shaded region in Fig.~\ref{fig:chi}(a). Component $\Pi_{33}^{ssc} = 0$ also implies $\Pi_{11}^{ssc} = 0$ \er{Pi_ssc}. However, the off-diagonal component $\Pi_{12}^{ssc}$ still survives. If we look at the form of $\chi_{11}$ in Eq.~\er{susc1}, we find that the pole in the $\mathcal{D}_{11}$ component is probed in ESR/EDSR by $\Pi_{12}^{ssc}$ which is non-zero even when $\Delta_Z=0$. It is because of this reason the pole in $\mathcal{D}_{11}$ also shows up in $\chi_{11}$ and, therefore, is ESR/EDSR active, as is evident by sharp red peak above the continuum in Fig.~\ref{fig:chi}(a). The red shaded region is the single-particle continuum of the in-plane sector, the edges of which coincides with that of the out-of-plane sector. This can be verified from the pole structure of single-particle bubbles, $\Pi_{ij}^{ss}$ and $\Pi_{ij}^{ssc}$, contributing to the spin-susceptibility.

\subsection{Spin response with magnetic field}
\label{ESR_finite}
We now discuss the effect of finite magnetic fields on ESR/EDSR. As discussed in Sec.~\ref{sec:se_Z}, the continuum becomes gapped at finite $\Delta_Z$. This supports collective modes both below and above the continuum as discussed in Sec.~\ref{finite_field}. At finite $\Delta_Z$, in addition to modes in the in-plane sector, the out-of-plane sector modes are also renormalized due to interaction. Moreover, all the components in Eq.~\er{Pi_ssc} are now non-zero which clearly indicates that whichever mode that showed up in Im$\mathcal{D}_{ij}$ will also show up in the imaginary part of $\chi_{ij}$, according to Eq.~\er{susc1}. Indeed, it can be seen in Fig.~\ref{fig:chi}(b) that all the modes below and above the continuum are excited at finite $\Delta_Z$. The parameter values for this figure has been chosen to be same as that in Fig.~\ref{fig:Z}(b), including the damping parameter. It can be checked, and also verified from the definition \er{susc1}, that the location of modes coincides in both these figures, which must be the case if these modes are probed in ESR/EDSR. 
To clarify, one of the red peaks in Fig.~\ref{fig:chi}(b) which is closest to the continuum above it is not well resolved as compared to that in Fig.~\ref{fig:Z}(b). We emphasize that the location of the peak in both these figures is the same, it's just the large spectral weight of the ESR/EDSR continuum (due to additional contribution from $\Pi_{11}^{ss}$, $\Pi_{11}^{ssc}$ and $\Pi_{12}^{ssc}$, as clear from the definition in Eq.~\er{susc1}) in Fig.~\ref{fig:chi}(b) makes it difficult to distinguish.

\section{Discussion and Conclusion}
\label{dis}
In this paper we have demonstrated the emergence of new collective modes in polar metals near their polar transition resulting from spin-orbit mediated interaction between electrons and soft phonons. We have identified the Zeeman energy as a tuning knob that can control the number, the energies and the character, namely electronic or phononic, of these collective modes. 

We have also shown that these emergent collective modes can be probed in optical absorption experiments. Both the phonon (Fig. \ref{zero_field}, Fig. \ref{fig:Z}) and the electronic (Fig. \ref{fig:chi}) responses bear fingerprints of the collective modes. Note that the electronic response is specific to the polar phase. It comes from both spin-susceptibility (electronic spin resonance) and optical conductivity (electronic dipole spin resonance), which are related to one another in the polar phase as explained in Sec. \ref{RPA}. In the non-polar phase both of these contributions vanish \cite{kumar_polar}.

We note that our calculations also apply for polar metals close to thermally driven polar transitions, if they are not strongly first order. This further broadens the range of possible candidate materials. Moreover, it indicates that for metals close to a second-order polar phase transition the relevant magnetic field scale should go to zero at the transition, allowing for experiments at low fields. However, we note that calculations presented here are only applicable away from the transition because the perturbative approach breaks down in its vicinity, see Appendix B of Ref.~\cite{kumar_polar} for detailed discussion of this issue. Finally, using magnetic doping instead of external magnetic field may raise the Zeeman splitting scale, such that the effects we predicted can be observed at larger frequencies and without the need to apply magnetic field.

A natural candidate for the realization of our proposal would be the polar Ca-substituted \cite{Wang2019} or oxygen-isotope substituted \cite{Itoh1999} SrTiO$_3$ which otherwise is a quantum paraelectric \cite{mueller, kamrannat, Collignon:2019}; electron doping in the polar phase of SrTiO$_3$ makes the system a polar metal. Another material of interest is KTaO$_3$ \cite{calvi:1995,Venditti_2023} that has a considerably larger spin-orbit coupling than SrTiO$_3$.

We now provide an estimate of the magnetic field strength that is required to change the number and character of the collective modes presented in Fig.~\ref{fig:Z}. We note first that our results predict at least one collective mode at low energies for arbitrarily nonzero $\Delta_Z$. 
However, to change the mode character and their number, the characteristic value of $\Delta_Z$ is of the order of the value of polar mode energy in the weak coupling limit (see Eq. \eqref{cond3}, where additionally $\omega_\parallel \sim \omega_\perp$).
Raman measurements give this information for the Ca-substituted doped SrTiO$_3$ (Sr$_{1-x}$Ca$_x$TiO$_{3-\delta}$)\cite{behnia_nature}. 
The classical polar metallic transition, associated with the softening of polar mode frequency, occurs there at roughly 20 K. The minimal mode frequency observed experimentally around that temperature is $\omega_\parallel \approx 12 \text{cm}^{-1} \approx 1.5$ meV, while at lower temperature it remains below  $20 \text{cm}^{-1} \approx 2.5$ meV. 
The corresponding magnetic field strength is, assuming the Landé $g$-factor for electrons to be 2 (see Ref.~\cite{kumar_polar}, and references therein, for detailed prediction for experiment), is roughly 17 T.

The polar metallic phase of KTaO$_3$  has not been yet realized experimentally. 
However, a polar transition has been observed in polycrystalline KTaO$_3$ thin films \cite{skoromets:2011} at 60 K. Slightly away from the critical point, the associated soft mode frequency is $\omega_\parallel \approx 7$ meV. If the polar transition survives in the metallic phase too (similarly as in SrTiO$_3$), the estimate for the magnetic field strength comes out to be roughly 77 T in KTaO$_3$ for the same ratio $\Delta_Z/\omega_\parallel = 1$. 
We note that depending on the strength of the electron-phonon coupling, effect of mode hybridization may appear at lower field. E.g., for the parameters used in Fig. \ref{fig:Z}, the mode number changes already at $\Delta_Z/\omega_\parallel = 0.8$, corresponding to field around 62 T. 

As shown in our previous work \cite{kumar_polar}, spectroscopy of collective modes can be used to deduce the value of the spin-orbit mediated electron-phonon coupling \er{coupling}. A similar procedure can be developed for deducing the coupling from ESR or EDSR experiments in the polar phase using the present results. In the polar phase, an alternative way to measure the coupling constant is by measuring the beatings of Shubnikov-de Haas (ShdH) oscillations in the polar phase arising due to spin splitting. From these beatings, one can estimate the static Rashba spin-orbit strength ($\alpha$) which is eventually related to the coupling constant $\lambda$, according to Eq.~\er{alpha}.


We note that the above estimates for SrTiO$_3$ and KTaO$_3$ are performed for the vicinity of the classical polar transition. It is possible that by tuning a parameter one can decrease the critical temperature down to the lowest value and reach the quantum limit; see Ref.~\cite{Wang2019} for SrTiO$_3$ where the quantum critical point is tuned by increasing doping density. 
However, spectroscopic data on the mode frequencies is not yet available. The predictions of this work should also apply in the vicinity (but not too close) to the polar QCP.



\section{Acknowledgements}
We thank P. Coleman, A. Damascelli, A. Klein, S. Maiti, D. Maslov and J. Ruhman for stimulating discussions.  P. C. is supported by DOE Basic Energy Sciences grant DE- SC0020353, as was A.K. during his time at Rutgers when this project was initiated.  P.A.V. was supported by a Rutgers Center for Material Theory Postdoctoral Fellowship during while he was at Rutgers. P.A.V. and P. C. acknowledge the Aspen Center for Physics where part of this work was performed, which is supported by National Science Foundation grant PHY-1607611. This work was partially supported by a grant from the Simons Foundation (P.A.V.).

\appendix
\section{Coherence factors for various correlation functions: spin current-spin current ($f_{ij}^{r\bar{r}, scsc}$), spin-spin ($f_{ij}^{r\bar{r}, ss}$), spin-spin current ($f_{ij}^{r\bar{r}, ssc}$) and spin current-spin ($f_{ij}^{r\bar{r}, scs}$)}
\label{appen:coh}
In this section we will provide explicit forms of coherence factors for all kinds of bubbles. Using the definition of spin-spin \er{ss}, spin-spin current \er{ssc}, spin current-spin \er{scs} and spin current-spin current \er{scsc} correlation functions, along with that of the Green's function \er{gr} with $\Delta_\bk$ given by Eq.~\er{es}, we calculate the corresponding coherence factors. More precisely, they are given by
\beq
\label{def:coh}
\begin{split}
f_{ij}^{r\bar{r}, ss} &= \text{Tr} \big[ \hat{\sigma}_i \hat{D}_{\bar{r}}(\bk) \hat{\sigma}_j \hat{D}_r(\bk) \big], \\
f_{ij}^{r\bar{r}, ssc} &= \text{Tr} \big[ \hat{\sigma}_i \hat{D}_{\bar{r}}(\bk) (\bk \times \hat{\bs})_j \hat{D}_r(\bk) \big], \\
f_{ij}^{r\bar{r}, scs} &= \text{Tr} \big[ (\bk \times \hat{\bs})_i \hat{D}_{\bar{r}}(\bk) \hat{\sigma}_j \hat{D}_r(\bk) \big], \\
f_{ij}^{r\bar{r}, scsc} &= \text{Tr} \big[ (\bk \times \hat{\bs})_i \hat{D}_{\bar{r}}(\bk) (\bk \times \hat{\bs})_j \hat{D}_r(\bk) \big],
\end{split}
\eeq
where $\hat{D}_s(\bk)$ is only the matrix part in the full definition of the electron Green's function as defined in the second line of \er{gr}. Here, $\Delta_\bk$ is given in Eq.~\er{es}. The chiral Green's function, $g_s(i\ve_m, \bk)$, does not have any contribute to coherence factors; they are instead used for frequency summation which we will show in Eq.~\er{freq_sum} of Appendix~\ref{appen:bubble}. We note that in this work we have assumed the momentum transfer to be zero ($q=0$); therefore, the $\bq$-dependence in $\hat{D}_s(\bk)$ has been suppressed explicitly. Also, in general, coherence factors are functions of polar and azimuthal angles: $f_{ij}^{r\bar{r}, ab}(\theta, \phi)$, where $a$ and $b$ could be either spin or spin-current. For brevity, we omit this dependence and write it simply as $f_{ij}^{r\bar{r}, ab}$ \er{def:coh}.

\paragraph{\textbf{Explicit form of $f_{ij}^{r\bar{r}, ss}$}}
\label{app_scsc}
Charge is conserved at $q=0$. So, all the charge components of the bubble are zero. However, spins are no longer a conserved quantity even at $q=0$. Therefore, the components of bubble with $(i, j) \in (1...3)$ are expected to survive. However, as discussed in Sec.~\ref{RPA} of the main text (see the text below Eq.~\er{compact}), because of $x \to -x$ and $y \to -y$ symmetry in the system, even some of the spin-sector components of the bubble vanishes. Therefore, we provide the expression of only those components of spin-spin coherence factor whose corresponding bubble is finite: 
\beq
\label{coh_ss}
\begin{split}
f_{11}^{r\bar{r}, ss} &= \frac{1}{2} \bigg[ 1 - r\bar{r} \frac{\Delta_Z^2}{\Delta_\bk^2} \bigg], \\
f_{22}^{r\bar{r}, ss} &= f_{11}^{r\bar{r}, ss}, \\
f_{33}^{r\bar{r}, ss} &= \frac{1}{2} \bigg[ 1 + r\bar{r} \frac{\Delta_Z^2}{\Delta_\bk^2} - r\bar{r} \frac{4\alpha^2k^2 \sin^2\theta}{\Delta_\bk^2} \bigg], \\
f_{12}^{r\bar{r}, ss} &= \frac{1}{2} i(r - \bar{r})\frac{\Delta_Z}{\Delta_\bk}, \\
f_{21}^{r\bar{r}, ss} &= -f_{12}^{r\bar{r}, ss}.
\end{split}
\eeq

\paragraph{\textbf{Explicit form of $f_{ij}^{r\bar{r}, ssc}$}}
Using the same argument as above, the coherence factors for only non-zero components of the spin-spin current bubble can be calculated:
\beq
\label{coh_ssc}
\begin{split}
f_{11}^{r\bar{r}, ssc} &= r\bar{r} \frac{\Delta_Z (2\alpha k \sin\theta)}{\Delta_\bk^2} k_y \sin\phi, \\
f_{22}^{r\bar{r}, ssc} &= r\bar{r} \frac{\Delta_Z (2\alpha k \sin\theta)}{\Delta_\bk^2} k_x \cos\phi, \\
f_{33}^{r\bar{r}, ssc} &= -r\bar{r} \frac{\Delta_Z (2\alpha k \sin\theta)}{\Delta_\bk^2} (k_x \cos\phi + k_y \sin\phi), \\
f_{12}^{r\bar{r}, ssc} &= -i(r - \bar{r})\frac{\alpha k \sin\theta}{\Delta_\bk} k_x \cos\phi, \\
f_{21}^{r\bar{r}, ssc} &= i(r - \bar{r})\frac{\alpha k \sin\theta}{\Delta_\bk} k_y \sin\phi.
\end{split}
\eeq
Although $f_{11}^{r\bar{r}, ssc}$ and $f_{22}^{r\bar{r}, ssc}$ are different, from the rotational symmetry in the x-y plane the corresponding bubbles $\Pi_{11}^{ssc}$ and $\Pi_{22}^{ssc}$ will be equal, which is indeed the case as presented in Eq.~\er{Pi_ssc} of the main text. Same way it is also easy to see that $f_{21}^{r\bar{r}, ssc} = -f_{12}^{r\bar{r}, ssc}$.

\paragraph{\textbf{Explicit form of $f_{ij}^{r\bar{r}, scs}$}}
The coherence factors for non-zero components of the spin current-spin bubble are
\beq
\label{coh_scs}
\begin{split}
f_{11}^{r\bar{r}, scs} &= r\bar{r} \frac{\Delta_Z (2\alpha k \sin\theta)}{\Delta_\bk^2} k_y \sin\phi, \\
f_{22}^{r\bar{r}, scs} &= r\bar{r} \frac{\Delta_Z (2\alpha k \sin\theta)}{\Delta_\bk^2} k_x \cos\phi, \\
f_{33}^{r\bar{r}, scs} &= -r\bar{r} \frac{\Delta_Z (2\alpha k \sin\theta)}{\Delta_\bk^2} (k_x \cos\phi + k_y \sin\phi), \\
f_{12}^{r\bar{r}, scs} &= -i(r - \bar{r})\frac{\alpha k \sin\theta}{\Delta_\bk} k_y \sin\phi, \\
f_{21}^{r\bar{r}, scs} &= i(r - \bar{r})\frac{\alpha k \sin\theta}{\Delta_\bk} k_x \cos\phi.
\end{split}
\eeq
One can notice from the above equation that all the diagonal components of $f_{ij}^{r\bar{r}, scs}$ are equal to those of $f_{ij}^{r\bar{r}, ssc}$. Moreover, $f_{12}^{r\bar{r}, scs} = -f_{21}^{r\bar{r}, ssc}$ and $f_{21}^{r\bar{r}, scs} = -f_{12}^{r\bar{r}, ssc}$.

\paragraph{\textbf{Explicit form of $f_{ij}^{r\bar{r}, scsc}$}}
Finally, the coherence factors for non-zero components of the spin current-spin current bubble can be calculated:
\beq
\label{coh_scsc}
\begin{split}
f_{11}^{r\bar{r}, scsc} &= \frac{1}{2} \bigg[ k_y^2 \bigg( 1 + r\bar{r} \frac{\Delta_Z^2}{\Delta_\bk^2} \bigg) - r\bar{r} k_y^2 \frac{4\alpha^2 k^2 \sin^2\theta}{\Delta_\bk^2} \\
&\hspace{2.5cm} + k_z^2 \bigg( 1 - r\bar{r} \frac{\Delta_Z^2}{\Delta_\bk^2} \bigg) \bigg], \\
f_{22}^{r\bar{r}, scsc} &= \frac{1}{2} \bigg[ k_x^2 \bigg( 1 + r\bar{r} \frac{\Delta_Z^2}{\Delta_\bk^2} \bigg) - r\bar{r} k_x^2 \frac{4\alpha^2 k^2 \sin^2\theta}{\Delta_\bk^2} \\
&\hspace{2.5cm} + k_z^2 \bigg( 1 - r\bar{r} \frac{\Delta_Z^2}{\Delta_\bk^2} \bigg) \bigg], \\
f_{33}^{r\bar{r}, scsc} &= \frac{1}{2} \bigg[ (k_x^2 + k_y^2) \bigg( 1 - r\bar{r} \frac{\Delta_Z^2}{\Delta_\bk^2} \bigg) \\
&\hspace{1cm} + r\bar{r} (k_x^2 - k_y^2) \cos2\phi \frac{4\alpha^2 k^2 \sin^2\theta}{\Delta_\bk^2} \\
&\hspace{1cm} + r\bar{r} 2k_xk_y \sin2\phi \frac{4\alpha^2 k^2 \sin^2\theta}{\Delta_\bk^2} \bigg], \\
f_{12}^{r\bar{r}, scsc} &= \frac{1}{2} i(r - \bar{r}) k_z^2 \frac{\Delta_Z}{\Delta_\bk}, \\
f_{21}^{r\bar{r}, scsc} &= -f_{12}^{r\bar{r}, scsc}.
\end{split}
\eeq

\section{Calculation of bubble}
\label{appen:bubble}
The functional form of all kinds of bubbles is same as that in Eq.~\er{compact} of the main text, except that the coherence factor, as discussed in Appendix~\ref{appen:coh}, is replaced by those of bubbles that we are calculating. In this section, we provide steps of the calculation of frequency summation and $k$-integral that show up in Eq.~\er{compact}. The frequency summation yields
\beq
\label{freq_sum}
\begin{split}
\Pi_{ij}^{ab}(i\Omega_n) &= \lambda^2 \int_K \sum_{r \bar{r}} f_{ij}^{r\bar{r}, ab} g_r (i(\omega_m + \Omega_n), \bk) g_{\bar{r}} (i\omega_m, \bk) \\
&= \lambda^2 \sum_{r\bar{r}} \int \frac{d^3k}{(2\pi)^3} f_{ij}^{r\bar{r}, ab} \frac{n_F(\ve_\bk^{\bar{r}} - \mu) - n_F(\ve_\bk^r - \mu)}{i\Omega_n - \ve_\bk^r + \ve_\bk^{\bar{r}}},
\end{split}
\eeq
where $\ve_\bk^s$ is the eigenvalue as given by Eq.~\er{es} of the main text.

The next step is the calculation of $\bk$-integral. We assume that the band-splitting caused by the combination of Rashba SOC and Zeeman field, $\Delta_\bk$ \er{es}, is small compared to the Fermi energy. This allows us to expand the Fermi function: $n_F(\ve_\bk^s - \mu) \approx n_F(\ve_\bk - \mu) + s(\Delta_\bk/2) n_F'(\ve_\bk - \mu)$, where $\ve_\bk = k^2/2m_b$. At $T=0$, $n_F(\ve_\bk - \mu) \approx \Theta(\mu - \ve_\bk)$ and $n_F'(\ve_\bk - \mu) \approx -\delta(\ve_\bk - \mu)$. This results in Eq.~\er{freq_sum} as
\beq
\Pi_{ij}^{ab}(i\Omega_n) = \frac{\lambda^2}{2} \sum_{r\bar{r}} \int \frac{d^3k}{(2\pi)^3} f_{ij}^{r\bar{r}, ab} \frac{(r-\bar{r}) \Delta_\bk \delta(\ve_\bk - \mu)}{i\Omega_n - (r-\bar{r}) \Delta_\bk/2}.
\eeq
As we see, the $\delta$-function immediately projects $k$ onto $k_F$ which is valid in the limit of large $\mu$. Rest is to do angle-integration which results in a log. The origin of log is the effective two-dimensionality of the $\bk$-integral as shown in Eq.~\er{demo2} of the main text.

\bibliography{referenceFile.bib}

\end{document}